\documentclass[article]{aa}
\usepackage[dvips]{graphicx}
\usepackage{txfonts}

\def\HII{H{\sc ii} }

\def\kms{\mbox{km~s$^{-1}$}}

\def\jff{\mbox{$J$=5$\rightarrow$4}}
\def\jdu{\mbox{$J$=2$\rightarrow$1}}
\def\juz{\mbox{$J$=1$\rightarrow$0}}

\def\iras{IRAS~21391+5802}

\begin{document}

\title{The stellar population and complex structure of the bright-rimmed cloud IC 1396N}

\author{M.\ T.\ Beltr\'an \inst{1} \and F.\ Massi \inst{2} \and R.\ L\'opez \inst{3} \and J.\ M.\
Girart \inst{4} \and R.\ Estalella \inst{3}} 

\offprints{M. T. Beltr\'an, \email{mbeltran@am.ub.es}}

\institute{Universitat de Barcelona, Departament d'Astronomia i Meteorologia, Unitat Associada
a CSIC, Mart{\'\i} i Franqu\`es 1, 08028 Barcelona, Catalunya, Spain  
\and
INAF-Osservatorio Astrofisico di Arcetri, Largo E.\ Fermi 5, 50125 Firenze,
Italy
\and
Departament d'Astronomia i Meteorologia, Universitat de Barcelona, Mart{\'\i} i Franqu\`es 1,
08028 Barcelona, Catalunya, Spain 
\and
Institut de Ci\`encies de l'Espai (CSIC-IEEC), Campus UAB, Facultat de Ci\`encies, Torre C-5,
08193, Bellaterra, Catalunya, Spain
}

\date{Received date; accepted date}

\titlerunning{IC 1396N}
\authorrunning{Beltr\'an et al.}

\abstract
{IC~1396N is a bright-rimmed cloud associated with an intermediate-mass star-forming region, where
a number of Herbig-Haro objects, H$_2$ jet-like features, CO molecular outflows, and millimeter compact
sources have been observed.}
{To study in detail the complex structure of the IC~1396N core and the
molecular outflows detected in the region and to reveal the presence of additional YSOs inside
this globule.}
{We carried out a deep survey of the IC~1396N region in the $J, H, K'$ broadband
filters and deep high-angular resolution observations in the H$_2$ narrowband filter with NICS
at the TNG telescope. The completeness limits in the 2MASS standard
are $K_{s}\sim$ 17.5, $H\sim$ 18.5 and $J\sim$ 19.5. }
{A total of 736 sources have been detected in all three bands within the area where the $JHK'$ images
overlap. There are 128 sources detected only in $HK'$, 67 detected only in $K'$, and 79 detected
only in $JH$. We found only few objects exhibiting a Near-Infrared excess and no clear signs of clustering
of sources towards the southern rim. 
In case of triggered star formation in the southern rim of the globule, this
could be very recent, because it is not evidenced through Near-Infrared imaging
alone. The H$_2$ emission is complex and knotty and shows a large number of molecular hydrogen features
spread over the region, testifying a recent star-formation activity throughout the
whole globule. This emission is resolved into several chains or groups of knots that
sometimes show a jet-like morphology. The shocked cloudlet model scenario previously proposed to
explain the V-shaped morphology of the CO molecular outflow powered by the intermediate-mass YSO
BIMA~2 seems to be confirmed by the presence
of H$_2$ emission at the position of the deflecting western clump. New possible flows have been
discovered in the globule, and some of them could be very long. In particular, the YSO BIMA~3 could be 
powering an old and poor collimated outflow.}
{}

\maketitle

\section{Introduction}

Bright-rimmed clouds (BRCs) found in \HII regions are potential sites of
triggered star formation due to compression by ionization/shock fronts.  Many of
them are associated with IRAS point sources with cold color indices (low dust
temperature), which are most probably Young Stellar Objects (YSOs) or
protostars. Such clouds are of deep interest from the point of view of ongoing
star formation. They frequently contain a small cluster of Near-Infrared (NIR)  stars that is
elongated toward the bright-rim tip or the ionizing star(s) of the \HII region
with the IRAS sources situated near the other end. There is a tendency for bluer
(i.e., older) stars to be located closer to the ionizing star(s), and for redder
(i.e., younger) stars to be closer to the IRAS sources. This asymmetric
distribution of the cluster members strongly suggests small-scale sequential
star formation or propagation of star formation from the side of the ionizing
star(s) to the IRAS position in a few times $10^5$ yr, as a result of the
advance of the shock caused by the UV radiation from the ionizing star(s)
(Sugitani et al.\ \cite{sugitani95}). Thus, BRCs represent one of the best laboratories for
studying  the star-formation process at different evolutionary stages.

A good example of BRC with ongoing star-formation activity is IC~1396N (BRC38; Sugitani et
al.\ \cite{sugitani91}), located in the Cep OB2 association at a distance of 750~pc (Matthews
\cite{matthews79}), and exposed to UV radiation from the O6.5 star HD 206267. The region is
associated with IRAS~21391+5802, a very young intermediate-mass object with a luminosity of
235~$L_\odot$ (Saraceno et al.\ \cite{saraceno96}), which is powering an extended CO bipolar
outflow (Sugitani et al.\ 1989). Beltr\'an et al.\ (\cite{beltran02}) have resolved the
millimeter emission towards IRAS~21391+5802 into an intermediate-mass source named BIMA~2
surrounded by two less massive and smaller objects, BIMA~1 and BIMA~3. Recent higher angular
resolution millimeter interferometric observations have revealed that the intermediate-mass
protostar BIMA~2 consists in fact of multiple compact sources (Neri et al.\ \cite{neri07}).  
The gas emission surrounding IRAS 21391+5802 traces different molecular outflows (Codella et
al.\ \cite{codella01}; Beltr\'an et al.\ \cite{beltran02}, \cite{beltran04}), some of them
possibly being powered by  yet undetected YSOs (Beltr\'an et al.\ \cite{beltran04}). Beltr\'an
et al.\ (\cite{beltran02}) have conducted a detailed study of the bipolar outflow  associated
with the intermediate-mass protostar BIMA~2, and shown that its complex morphology and
kinematics are possibly the result of the interaction between the outflow and the dense cores
surrounding the protostar. NIR images of the region have also revealed the presence of a
number of small scale molecular hydrogen and Herbig-Haro (HH) flows (Nisini et al.\
\cite{nisini01}; Sugitani et al.\ \cite{sugitani02a}; Reipurth et al. \cite{reipurth03};
Caratti o Garatti et al.~\cite{caratti06}). This evidence for ongoing star-formation
activity at the head of the cometary globule together with the relatively proximity of the
region make IC~1396N one of the best candidates to study potential sequential star formation.

To do a complete and uniform census of the young stellar population in the globule and reveal
the presence of additional young sources inside the globule, deep NIR observations at $J,
H$ and $K'$ have been carried out. In addition, deep high angular resolution observations in
the S(1) v=1--0 line of H$_{2}$ at 2.12~$\mu$m have also been performed to investigate the 
complex structure of
this globule, and the morphology of the shocked gas that traces the interaction between the
outflows in the region and the dense clumps surrounding the YSOs. The results of this NIR
study are presented here.

\section{Observations and data reduction}
\label{obs}

The images were taken with NICS (Baffa et al.~\cite{baffa01}) at the 3.58-m Telescopio
Nazionale Galileo (TNG) telescope (ORM, La Palma, Spain) through the standard $J,H,K'$ 
broadband filters and the H$_{2}$ narrowband filter centered at 2.12~$\mu$m, during the nights between 16--17 July
2005. The plate scale is $0.25\arcsec$/pixel, yielding a field of view of $\sim$ $4.2 \times
4.2$ arcmin$^{2}$. Both in $K'$ and in H$_{2}$, two positions roughly $100\arcsec$ apart
(east-west) were imaged, such as to have an overlapping field, $\sim$ $150\arcsec$ wide (in
RA), enclosing the globule. In $H$, the two imaged positions are separated by
$\sim$ $50\arcsec$ east-west, so the overlapping field is $\sim$ $200\arcsec$ wide. Due to
shortage of time, only one field could be  imaged in $J$, centered on the globule. The seeing
was $\sim$ $0.8\arcsec$ in $K'$, $H$ and H$_{2}$, and $\sim$ $0.9\arcsec$ in $J$.  In $K'$ and
$H$, groups of five on-source integrations with a dithering of up to $10\arcsec$ in RA and DEC
were interspersed between groups of five off-source integrations. The off-source fields are
located $\sim$ $6\arcmin$ from the target and had been chosen through examination of 2MASS
images. The dithering of the off-source frames is up to $20\arcsec$ in RA and DEC. In $J$,
groups of two on-source images were interspersed between groups of two off-source images.
Ditherings and off-source fields are the same as above.  Finally, in H$_{2}$ pairs of one
on-source and one off-source images were taken, with the  same ditherings and off-source
fields as above. Each frame was integrated 5 or 10~s in $K'$, depending on the background
level; the total integration time is 600~s for each of the two positions. At $H$, each
individual integration  is 20~s and the total integration time is 600~s for each of the
two positions, as well. At $J$, each individual integration is 100~s and the total
integration time is 600~s. In H$_{2}$, the individual integration times are 100
or 150~s,
depending on the background level, and the total integration time is 2700~s for each of the
two positions.  

\begin{figure*}
   \centering
   \includegraphics[width=13cm,angle=0]{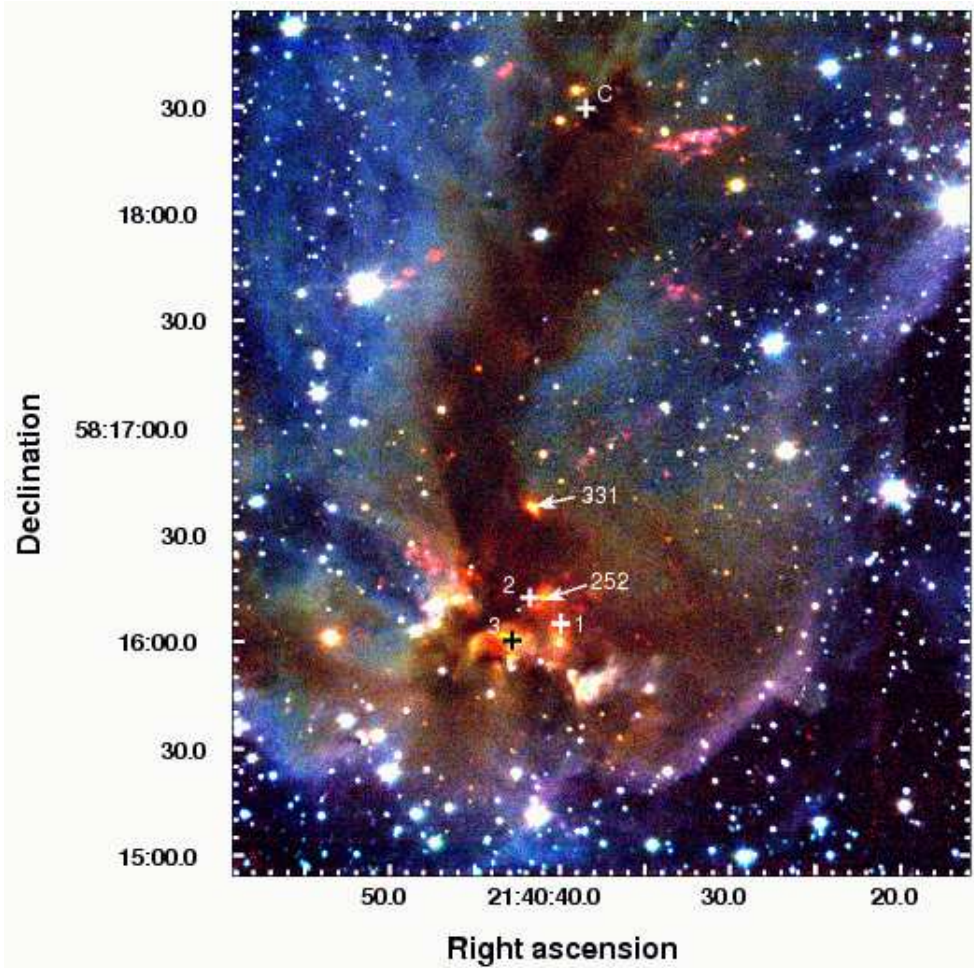}
      \caption{Three color composite image of IC~1396N ($J$, blue, $H$, green, $K'$, red) 
      taken with NICS at TNG. The 
      black and white crosses show the positions of the 3.1~mm sources, BIMA~1,
2, and 3 from Beltr\'an et al.~(\cite{beltran02}), while the white cross in the top shows the position of the 1.3~mm
continuum source C detected by Codella et al.~(\cite{codella01}). Also labeled are
         two Class I sources discussed in the text.         
	 \label{fig:kmap}
              }
   \end{figure*}

Each frame was first corrected for cross-talk using the routine provided on the TNG web page
(http://www.tng.iac.es). Data were then reduced in the standard way by using IRAF\footnote{IRAF
is distributed by the National Optical Astronomy Observatories, which are operated by the
Association of Universities for Research in Astronomy, Inc., under cooperative agreement with
the National Science Foundation.} routines. Flat-field frames were acquired at sunset.
Differential flat-field images were constructed for $K'$ and H$_{2}$, whereas all available
frames with roughly the same mean level of counts  were averaged together for $H$ and $J$. All
on-source and off-source frames were then flat-field corrected. Sky frames were constructed by
median-averaging the  six off-source frames closest to each on-source frame (generally, three
preceding and three following), after removal of the imaged  stars. The sky frames were then
subtracted from the corresponding on-source frames. At $K'$ and H$_{2}$, the sky-subtracted
images were multiplied by a factor when obtained  with different individual exposure times, such
as to ``convert'' the counts of  all frames in a same band to a same exposure time. After
bad-pixel correction, all images in a same band were registered and combined together by using a
median filter. The composite three color $JHK'$ image of the area where the $JHK'$ frames
overlap is shown in Fig.~\ref{fig:kmap}. 

Photometry on all mosaiced images was performed by using DAOPHOT (in IRAF). The detected stars
were retrieved by running DAOFIND and, subsequently, by a visual check in order to discard fake
detections and add undetected faint sources. Aperture photometry was carried out through PHOT, by
adopting an aperture $\sim$ $1 FWHM$ in radius and an annulus $\sim$ $2 FWHM$ wide with an inner
radius $\sim$ $2 FWHM$. The weather was barely photometric, so the calibration was performed by
cross-correlating the sources found in the $JHK'$ bands and the 2MASS point source catalog. A
linear relation in the $J$--$H$ or $H$--$K'$ instrumental colors was fitted to the found pairs of
instrumental magnitude and 2MASS magnitude and the corresponding instrumental color. Hence, the
given magnitudes are in the 2MASS system ($JHK_{s}$).  The color coefficient is always less than
0.06 in each band. As a check for consistency, we compared our $K_{s}$ photometry and that of
Nisini et al.\  (\cite{nisini01}) for the isolated sources out of those listed by those authors.
Our $K_{s}$ values are on average $0.34\pm0.28$ mag dimmer than those by Nisini et al.\
(\cite{nisini01}). This is very likely due to the much worse seeing ($\sim$ 2--3$\arcsec$) and the
much coarser sampling ($\sim$ $1\arcsec$) of the PSF in the data reported by Nisini et al.\
(\cite{nisini01}), given the highly variable background level in the region.

Additional photometry of the detected H$_{2}$ features was performed from the narrowband  H$_{2}$
image. The H$_{2}$ filter is centered at the 2.12~$\mu$m line of molecular hydrogen. Continuum 
emission falls within the bandpass, as well as line emission. Through photometry in H$_{2}$ and
$K'$, we estimated the fraction of stellar continuum affecting the total  counts in the H$_{2}$
frame.  First, based on the characteristics of the two filters,  two scale factors can be derived
by which we multiplied the H$_{2}$ and $K'$ images, and then we subtracted the latter from the
former. The resulting subtracted image contains only the line emission falling within the H$_{2}$
filter. The calibration was performed by carrying out stellar photometry on the H$_{2}$ original
image, the same way as for $JHK'$ but also including an aperture  correction. The retrieved stars
were cross-correlated with those found in the $HK'$ bands and their correct flux was derived by
interpolation with the corresponding ones at $H$ and $K_{s}$ in the 2MASS system. Through a fit,
we determined the conversion factor from counts to  flux. The detection limit (at a $3\sigma$
level) is $\sim$ $10^{-15}$ erg cm$^{-2}$ s$^{-1}$ arcsec$^{-2}$. The procedure outlined above
yielded an image where all stellar sources were efficiently removed, therefore only containing
H$_{2}$ line emission knots. We classified each local emission peak as a knot and defined a
polygon around each knot such as to include emission down to a $\sim$ $3\sigma$ limit.  Polygon
borders for close-by knots were chosen by eye, based on morphological criteria.  Photometry was
carried out by using POLYPHOT in IRAF.

The astrometric calibration was performed by deriving the positions of 14 relatively bright, isolated stars
spread all over the $K'$ frame, and correlating them with their 2MASS coordinates. By a fit (using
STSDAS routines in IRAF) we obtained the transformation between frame and equatorial coordinates,
allowing an accuracy of $\sim$ 1$\arcsec$.

\section{Results and discussion}

\subsection{The stellar population}
\label{population}

Within the area where the $JHK'$ images overlap (see Fig.~\ref{fig:kmap}), we found 736 sources
detected in all three bands, 128 sources detected only in $HK'$, 67 sources with a $K'$
detection only, and 79 sources detected only in $JH$. The sources with $HK'$ or $K'$ detections
only  are preferentially located towards the globule (see, e.\ g.\ Fig.~\ref{fig:clustmap}), as
expected for heavily extincted stars. Conversely, the sources with $JH$ detections tend to be
located outside the globule, indicating that these are just faint stars.

\begin{figure}
   \centering
   \includegraphics[width=7.5cm,angle=-90]{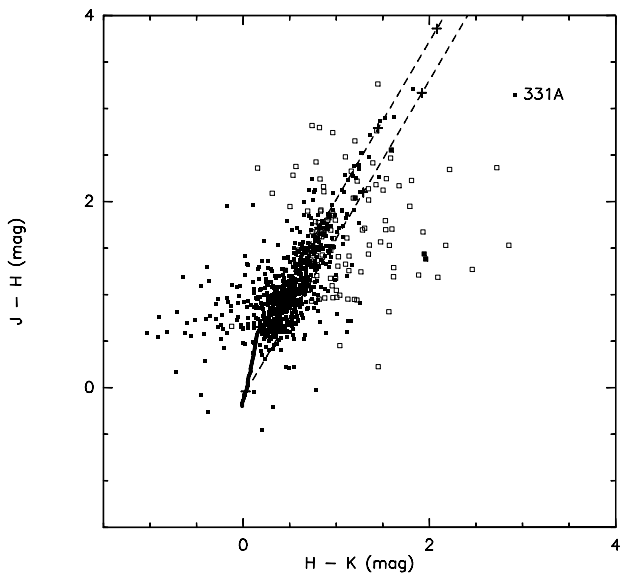}
      \caption{Color-color diagram of the NIR sources found
        within the area where $JHK'$ images overlap. Full squares
        are sources with detection in all bands, empty squares are
        sources with detection in $HK_{s}$ only (hence, the shown $J$--$H$ is 
        a lower limit). The solid line marks the main sequence
        (in the 2MASS system), the dashed lines follow the reddening law
        by Rieke \& Lebofsky (\cite{rieke85}) with crosses at intervals
        of $A_{v} = 10$ mag.
         \label{fig:col-col}
              }
   \end{figure}

\begin{figure}
   \centering
   \includegraphics[width=7.5cm,angle=-90]{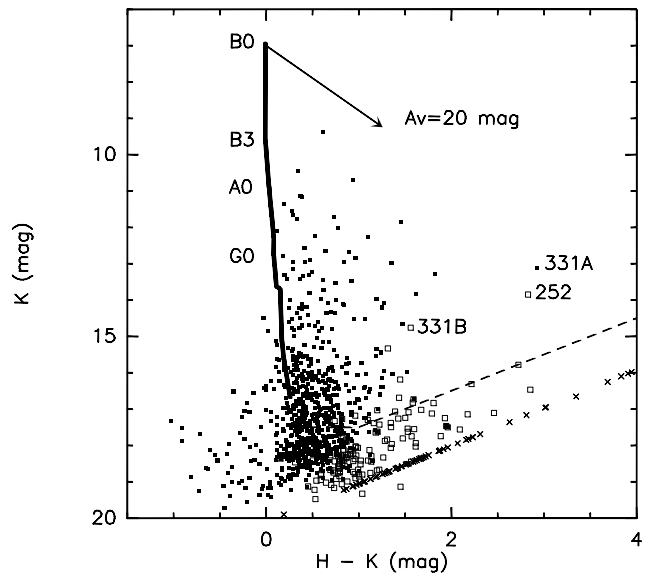}
      \caption{Color-magnitude diagram of the NIR sources found
        towards the area where $JHK'$ images overlap. Full squares
        are sources with detections in all bands, empty squares are
        sources with detections in $HK_{s}$ only and crosses are sources
        detected at $K_{s}$ only (hence, the shown $H$--$K_{s}$ is 
        a lower limit). The solid line marks the zero age main sequence
        (from Allen (\cite{allen76}) and Koornneef (\cite{koorn83}) 
        after conversion to the 2MASS system) at a distance of 750 pc, 
        the dashed lines indicate the completeness
        limit and an arrow is drawn showing a reddening $A_{V} = 20$ mag 
        according to the reddening law by Rieke \& Lebofsky (\cite{rieke85}).
        A few spectral types on the ZAMS are labeled. 
         \label{fig:mag-col}
              }
   \end{figure}

\begin{figure}
   \centering
   \includegraphics[width=8.7cm,angle=0]{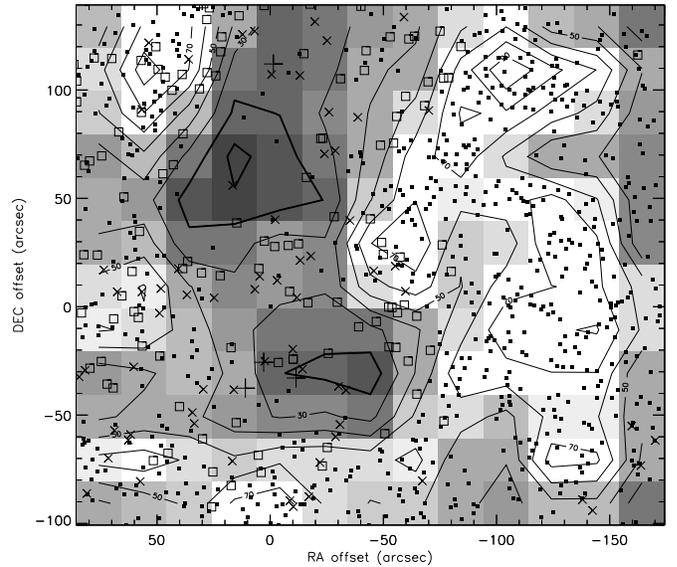}
      \caption{Stellar surface density map (in stars arcmin$^{-2}$)
        of all sources detected in the $K_{s}$ band up to $K_{s} = 20$.
        Contours range from 10 to 90~stars arcmin$^{-2}$ in steps of
        10 stars arcmin$^{-2}$.
        The contours at 10 and 20 stars arcmin$^{-2}$ are thicker. Grey scale 
	ranges from $-15$~stars arcmin$^{-2}$ (black) to 69~stars arcmin$^{-2}$ (white).
        The field shown has been imaged at all bands and the offsets
        are in arcsec from the location of source \# 331A (HH777 IRS).
        The positions of sources with detections in all bands are
        marked by filled squares, those of sources with detections 
        at $HK_{s}$ only by empty squares and those of sources with only
        a $K_{s}$ detection are marked by small crosses.
        The three large crosses around ($0\arcsec,-30\arcsec$) are the compact embedded
        sources BIMA 1, 2 and 3 detected by Beltr\'{a}n et al.\ (\cite{beltran02}) at
	millimeter wavelengths, and that at around ($0\arcsec, 110\arcsec$) is the 1.3~mm
        continuum source C detected by Codella et al.~(\cite{codella01}).
         \label{fig:clustmap}
              }
   \end{figure}

We obtained histograms of the number of sources as a function of magnitude
by binning the number of sources detected in all three bands in magnitude intervals. Then,
we adopted as completeness limit in each band the magnitude where the corresponding histogram peaks:
$K_{s}$$\sim$ 17.5, $H$$\sim$ 18.5 and $J$$\sim$ 19.5. When adding also the sources with detections
in two or one bands only, the peak does not shift in any of the histograms but $K_{s}$,
where it appears to move towards $K_{s}$$\sim$ 18. The derived completeness limits are roughly $1.5$
mag below our detection limits (at a $3\sigma$ level). 
An estimate of the minimum stellar mass detectable all over the globule
can be obtained by using pre-main sequence (PMS) 
evolutionary tracks. However, one has to assume an age and a maximum extinction
for the stellar population. As for the age, Lada \& Lada (2003) noted that
the embedded phase of star cluster evolution lasts 2--3~Myrs and clusters
older than 5 Myrs are rarely associated with molecular gas. This is in accord with the age
of the open star cluster Trumpler 37, surrounding the globule
($\sim$ $3\times10^{6}$ yrs; Getman et a.\ \cite{getman07}).  The age of the star exciting
the PDR around the globule may also give a hint of the age of the stellar
population, since this star either triggered star formation in the core, or began inhibiting it
by starting core destruction. 
If HD 206267 is an O6.5 V star (Walborn \& Panek \cite{wapa}), then its lifetime in the
main sequence is $\sim$ $6\times10^{6}$ yrs (e.\ g.\ Vanbeveren et al.\ \cite{vanbe}),
that roughly agrees with the times given above. On the other hand, the globule shows
clear signature of much younger stars and protostars, and the dynamical timescales estimated
from the jets (see Sect.~\ref{H2_emission}) are even shorter, $\sim$ $10^{3}$ yrs. 
We can therefore assume an age of $10^{6}$ yrs as a sort of upper limit, 
since younger low-mass PMS stars are  brighter and, then, more easily detectable
(lowering the mass detection limit).
As for the maximum extinction, Getman et al.\ (\cite{getman07}) quote a few authors
to conclude that the absorption through the core is $A_{V}$$\sim$ 9--10~mag.
Figures~\ref{fig:col-col} and \ref{fig:mag-col} clearly
show that this value is probably too low and most of the detected stars 
exhibit $A_{V} \leq 20$ mag. Nevertheless,
there are a few sources with $A_{V}$ up to $\sim$ 30~mag. More extincted sources
could be not represented in the diagrams just because too faint, thus biasing any estimates
based on the plots. In fact, towards BIMA~2, we can
derive an extinction much larger than $100$ mag from the continuum millimeter data. 
However, this is clearly a less evolved, very young region amounting to a small fraction
of the globule. We can probably assume that for most of the globule the extinction does not
exceed a canonical $A_{V}$$\sim$ 30--40~mag.  

We adopted the PMS evolutionary tracks 
by Palla \& Stahler (\cite{palla99}), along with the reddening law
by Rieke \& Lebofsky (\cite{rieke85}). Hence, assuming an age of $10^{6}$ yrs, PMS stars
of $\sim$ 0.1~$M_{\sun}$ are within the completeness limit at $K_{s}$ for $A_{V} = 30$ mag and
within the detection limit at $K_{s}$ for $A_{V} = 40$ mag. The same for PMS stars of
$\sim$ 0.4 $M_{\sun}$ at $H$, whereas PMS stars of $\sim$ 0.8~$M_{\sun}$ 
are within the detection limit at $J$  for $A_{V} = 30$~mag and
PMS stars of $\sim$ 2~$M_{\sun}$ are within the completeness limit at $J$ for
$A_{V} = 30$ mag. However, these magnitudes refer to ``naked'' stars, i. e., stars
without a circumstellar disk. 

%

%

The color-color diagram (CCD; $J$--$H$ vs.\ $H$--$K_{s}$) of the NIR sources found within the area
where $JHK'$ images overlap is shown in Fig.~\ref{fig:col-col}. The main sequence locus is also
drawn, by using the colors of Koornneef (\cite{koorn83}) after conversion to the 2MASS system
through the relations given by Carpenter (\cite{carpenter01}). The CCD is consistent with that
shown in Nisini et al.\ (\cite{nisini01}), in that most of the stars fall within the reddening
band of the main sequence and almost all those exhibiting a NIR excess lie only slightly below the
reddening band. The points spread around the main sequence with larger NIR  excesses are mostly
faint sources found at the edge of the images, then affected by large errors. However, our source
\# 331 (labeled in figure) actually exhibits a large NIR excess. This source coincides with source \# 8
in Nisini et al.~(\cite{nisini01}) and HH777 IRS in Reipurth et al.\ (\cite{reipurth03}).
According to the latter authors, this source could be binary because it seems to be powering two flows,
a major HH flow that expands towards the southwest, labeled HH777 by Reipurth et al.\
(\cite{reipurth03}), and a northwestern flow labeled G by Nisini et al.\ (\cite{nisini01}). The
source is neatly elongated with respect to the PSF of a single star, with the size of the major
axis twice that of the minor axis, which suggests the binarity of the source. The nearby stars do
not show such an elongation, therefore we discarded any possible focus effect. The elongation of
source \# 331 is very clear in the $H$ filter, in which it has been possible to deconvolve the
emission into two stars \# 331A and \# 331B by PSF-fit photometry with DAOPHOT in IRAF. In the
$K'$ filter the elongation is also evident, but the PSF-fit photometry appears to be less precise.
Nevertheless, we cannot rule out the possibility that \# 331B may be just radiation from \# 331A
scattered by dust through a cavity. This scenario would be consistent with the fact that the
elongation of source \# 331 roughly coincides with the direction of the southwestern HH777 flow (see Fig.~2 of
Reipurth et al.~\cite{reipurth03}). However, as seen in Figs.~\ref{fig:kmap} and
\ref{knots_total}, the other H$_2$ flow detected nearby, the northwestern flow labeled 
G, also points right back towards source \# 331, which suggests that its
powering source is also located at that position. Hence, based on the fact that there are two
outflows associated with this position, we favor the scenario of binarity to
explain the elongation of source \# 331. The $HK_{s}$
magnitude of \# 331A is within $0.4$ mag of those listed in the 2MASS catalog but,
whereas its colors are consistent with those given by Nisini et al.\ (\cite{nisini01}), the
$K_{s}$ value we derive is almost 1 mag larger than that measured by Nisini et al.\
(\cite{nisini01}). This is still consistent with the difference found between the two
photometries (see Sect.~\ref{obs}), also given that the source has been resolved into two
close-by companions that
appear to be embedded in a small
diffuse nebulosity that probably could not be resolved by Nisini et al.\ (\cite{nisini01}).
However, a degree of intrinsic variability cannot be excluded, as well.

The color-magnitude diagram (CMD; $H$--$K_{s}$ vs.\ $K_{s}$) is shown in Fig.~\ref{fig:mag-col}
for the NIR sources in the same area as above. As seen in this diagram, an upper limit
for the spectral type of the ZAMS stars in the cloud would be B1--B0, which corresponds to a
stellar mass of $\sim$ 17--20~$M_\odot$ (Vacca et al.~\cite{vacca}). Since such massive stars
evolve along the ZAMS from 8--10~$M_{\sun}$ on (e.\ g.\ Palla \& Stahler \cite{psone})
at the end of their accretion phase, this can be considered as a robust upper
limit for the mass of the stars associated with IC1396N, irrespective of their age.
Most of the points lie within
$A_{V} = 20$ mag of the zero age main sequence (ZAMS). However, source \# 252
has similar $K_{s}$ and $H$--$K_{s}$ as \# 331A, both objects lying farther from the ZAMS than
the remaining stellar population. Source \# 252 is also embedded in a patch of diffuse
nebulosity and is located near a cluster of H$_{2}$-emission blobs already identified
by Nisini et al.\ (\cite{nisini01}) as knot A, and the cluster of compact
radio sources found by Beltr\'{a}n et al.\ (\cite{beltran02}). These facts suggest
that sources \# 252 and \# 331A may be in a similar evolutionary stage, although 
this cannot be fully proved because
\# 252 has not been detected at $J$. They are located towards the center of
the globule, $\sim$ $26\arcsec$ apart (see Fig.~\ref{fig:kmap}). Source
\# 252 lies close to the IRAS uncertainty ellipse, north-west of it. 
Projecting them back onto the ZAMS in the CMD, along the reddening
vector, identify them as stars of spectral type B0 to B3. This has to be considered as
a ``lower'' limit for their actual spectral type (i.\ e., they are of later spectral type), 
since \# 331A exhibits a NIR excess
and probably also \# 252 does have it. Hence, based on the location of \# 331A in
the CCD, they might be Class I sources of intermediate mass (e.g.\ Sugitani et
al.~\cite{sugitani02b}).

Getman et al.~(\cite{getman07}) used CHANDRA X-ray observations of IC~1396N,
complemented with Spitzer/IRAC photometry and the available NIR photometry,
identifying 25 likely stellar members of the globule. Although all are
associated with IRAC MIR sources, 6 of them do not have a NIR counterpart either
in the  2MASS catalog or in the list of Nisini et  al.~(\cite{nisini01}). We
have found three new matches, e.\ g.\ sources 70, 76, and 80 (see Table~2 of
Getman et al.~(\cite{getman07})), corresponding to our sources 224
($K_{s}$$\sim$ 16), 223  ($K_{s}$$\sim$ 16) and 196 ($K_{s}$$\sim$ 17.7),
respectively. They are all undetected  both in $J$ and in $H$, confirming their
nature of heavily extincted source. The remaining 3 X-ray sources without a NIR
counterpart lie in the area of the BIMA sources and their X-ray spectra are
heavily absorbed ($N_{\rm H} \ga 10^{23}$ cm$^{-1}$). One of them (source 66)
has been proposed as X-ray counterpart of the protostar BIMA~2.   However, note
that our sources \# 331A (and 331B) and 252 do not have an X-ray counterpart in
the catalog of Getman et al.~(\cite{getman07}). These authors quote a
completeness limit (in mass) of $0.4$ $M_{\sun}$, higher than our completeness
limit in $K_{s}$ but similar to our completeness limit in $H$. We suspect that
their completeness limit may be even higher, since the reddening towards the
globule may be at least twice larger than adopted by them. It is noteworthy that
Getman et al.~(\cite{getman07}) found X-ray counterparts of possible Class I
sources with high absorption and no NIR counterparts, but fail to detect our
sources \# 331A and 252.  This would be consistent with \# 331A and 252 being
young intermediate-mass (proto-)stars. If \# 331 is actually a double system,
then its companion (possibly \# 331B) might either be another young
intermediate-mass star, or a low-mass protostar (with a mass below the X-ray
completeness limit). Moreover, sensitive millimeter interferometric
observations did not detect \# 252 (Beltr\'an et al.\
\cite{beltran02}; Neri et al.~\cite{neri07}), which rules out the presence of the expected massive 
enough circumstellar disk. Hence, further data are needed to clarify its nature and
its identification as an intermediate-mass Class I source remains highly speculative.

%

\begin{figure*}
\centerline{\includegraphics[angle=0,width=15cm]{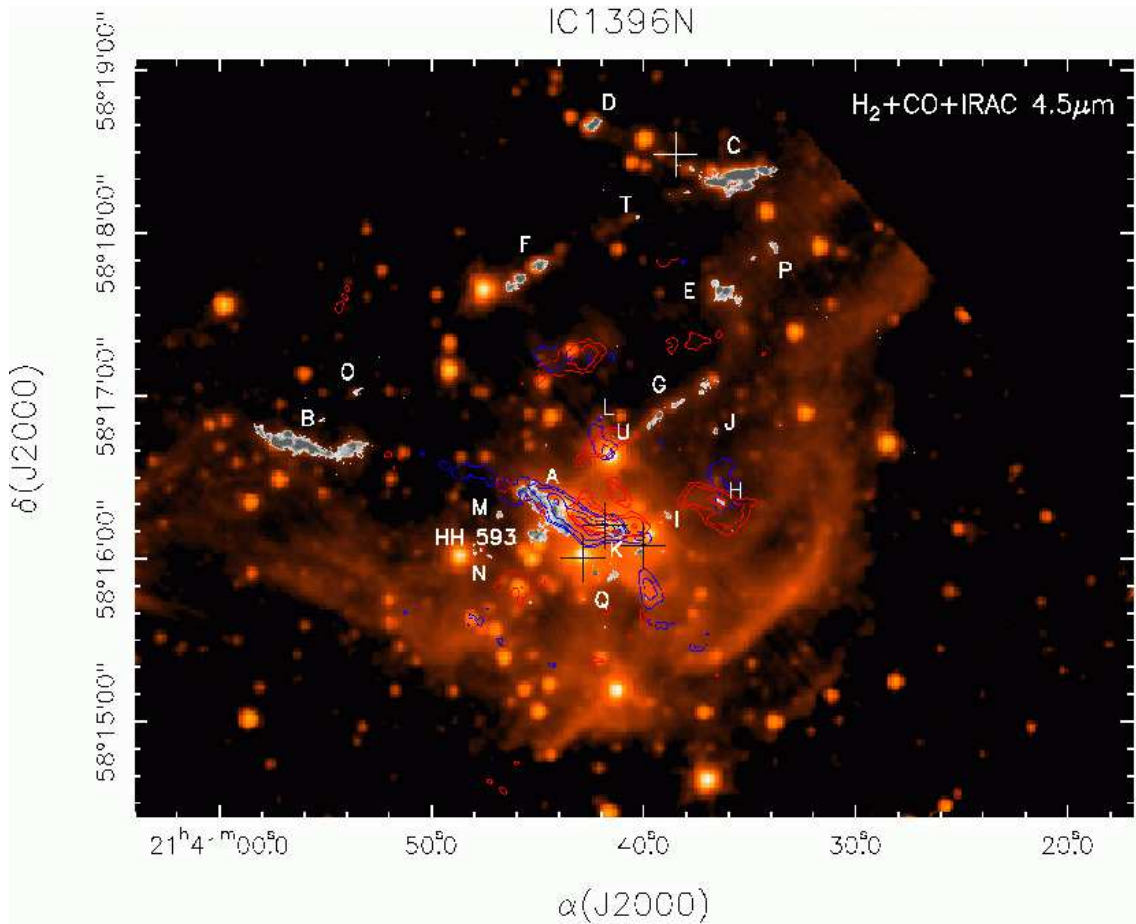}}
\caption{H$_2$ (2.12~$\mu$m) image (continuum-subtracted) in grey-scale and CO (\juz)
emission in red and blue contours (Beltr\'an et al.\ \cite{beltran02}) overlaid with an IRAC
4.5~$\mu$m image. The H$_2$ emission shows a large number of molecular hydrogen
 features, many already found by
Nisini et al.~(\cite{nisini01}), Reipurth et al.~(\cite{reipurth03}), and Caratti o
Garatti~(\cite{caratti06}), 
 whose nomenclature we use and expand. The black crosses show the positions of the 3.1~mm sources, BIMA~1,
2, and 3 from Beltr\'an et al.~(\cite{beltran02}), and the white one shows the position of the 1.3~mm
continuum source C detected by Codella et al.~(\cite{codella01}).}
\label{h2-spitzer}
\end{figure*}

\subsection{Triggered star formation?}

In the X-ray source population towards IC~1396N, Getman et al.~(\cite{getman07}) have found  a
clear clustering of sources at the southern rim, with an elongated spatial distribution, and an
evolutionary gradient (interpreted as an age  gradient), oriented towards the exciting star.
These authors interpret this geometric and age distribution in terms of triggered star
formation by passage of \HII region shocks into the molecular globule. We have searched for
evidence in the NIR of age gradients in the south-north direction or clustering of stars
towards the rim, but found none. The number of sources with evidence of NIR excess towards the
globule is too low, so any analysis of the stellar population in the NIR {\em alone} is bound
to remain inconclusive with respect to the identification of age gradients. Regarding the
geometric distribution of the sources in the NIR, there are no clear signs of clustering
towards the rim (even within the area where X-ray sources cluster), as shown by the  map of the star surface density (Fig.~\ref{fig:clustmap}),
which was obtained by  counting all sources with a detection at least in the $K_{s}$ band (up
to $K_{s} = 20$) in squares of $40 \arcsec \times 40 \arcsec$, displaced by $20 \arcsec$ both
in RA and in DEC. The number of sources  decreases in going from the southern edge of the
globule to the northern one; in the CCD, within the extinction band of the main sequence the
upper and lower limits of extinction initially increase, then decrease close to the northern
edge, as expected. What is more, the  $JHK_{s}$ sources lying below the main sequence reddening
band in the CCD tend to cluster out of the eastern and western edge of the globule, but much
less so towards the southern border.  As discussed above, our $JHK_{s}$ colors are biased
towards unextincted PMS stars of all masses and heavily extincted intermediate-mass young
stars. Instead, $HK_{s}$ colors only can evidence heavily extincted PMS stars down to $0.4$
$M_{\sun}$. These appear to be mostly located towards the globule, but it is still difficult to
evidence a significant clustering towards the southern border.  A clearer clustering of
reddened objects  occurs towards the center of the globule. This is also visible as an increase
of surface density (Fig.~\ref{fig:clustmap}) north of sources \# 331 (A and B),  mostly due to
sources with $HK_{s}$ or $K_{s}$ detection only. 

Therefore, only very few or no sources with a NIR excess clusters towards the southern 
rim, as far as extinction is low and we are sensitive to very low masses. Of course, 
a number of stars with a NIR excess may concentrate north of the southern edge of the 
globule, towards the BIMA sources, were extinction is much higher.
Then, any triggered YSOs may be still too young to be evidenced
through NIR imaging alone. This would be consistent with the star surface density map
(Fig.~\ref{fig:clustmap}), which shows  two surface density minima towards the globule, one in
the southern part and one in the northern part. The southern one is close to the position of
the embedded sources BIMA~1, BIMA~2, and BIMA~3 detected by Beltr\'{a}n et
al.~(\cite{beltran02}), and the northern one lies along the gas elongation visible in CO, CS
(Codella et al.~\cite{codella01}) and H$^{13}$CO${^+}$ (Sugitani et al.~\cite{sugitani02a}).
These two surface density minima outline the more extincted (densest)
parts of the globule, not yet visible at NIR wavelengths.

By decreasing the limiting magnitude to $K_{s} = 18$ (i. e., the completeness limit),  one
obtains a similar surface density map, but the ``bridge'' of sources crossing the two minima
almost disappears, confirming it mostly arises from the detection of faint sources.  If this
were a real group of young stars associated with the globule, then their birth could hardly be
explained as triggered, since they are located north of the group of the youngest protostars
(the three compact radio sources found by Beltr\'{a}n et al.~\cite{beltran02}), farther from
the ionization front. What is more, in going from south to north, one finds the cluster of
Class~0/I sources observed at millimeter wavelengths by Beltr\'an et al.~(\cite{beltran02})
and Neri et al.~(\cite{neri07}), then source \# 331A, which is definitely a more evolved
object, and then millimeter source~C (Codella et al.~\cite{codella01}), which is a 
deeply embedded and very young object. Therefore, from the NIR and millimeter observations it
is clear that not all the star formation in the globule can be explained in terms of
triggering.  

The difficulty in evidencing any stellar population associated with the globule based on
NIR photometry only was already noted by Getman et al.~(\cite{getman07}). As discussed,
X-ray observations proved much more efficient in selecting this population, although clearly
failing to probe all young stars and protostars. By combining X-ray and Spitzer/IRAC observations,
they could indeed find a number of Class~I sources towards the globule. This
confirms what we inferred from our NIR images. Also, those authors estimate that the total population
of T~Tauri stars of the globule is $\sim$ 30, also consistent with the lack of
a significant increase in the NIR source surface density towards the rim. 
Nevertheless, the lack 
of sources exhibiting a clear NIR excess close to the rim is unusual for a stellar population
of $\sim$ $10^{6}$ yrs old and strongly suggests that the intense UV radiation
may have affected their circumstellar environments, suddenly stopping their growth. This would
be confirmed by the low masses inferred for the counterparts of the X-ray sources associated
with the globule. Most of them have estimated masses between 0.2--0.5~$M_{\sun}$, and
they would be even less massive if they were younger than $10^{6}$ yrs, as assumed by
Getman et al.~(\cite{getman07}). Hence, we caution against interpreting an evolutionary
gradient as an age gradient in an environment like the edge of a bright rimmed cloud. 
The eroding ionization front,
in fact, may have dispersed the circumstellar environment of protostars downstream of it 
leaving them as naked stars, without significantly affect protostars upstream of it.
In this case, an evolutionary gradient may not correspond to a real age gradient.  
As shown, star formation activity is present throughout the whole core, even in the northern
part. Then, at the moment it appears really difficult proving that star formation in the southern rim
has been triggered.





\subsection{H$_2$ emission}
\label{H2_emission}

\begin{figure*}
\centerline{\includegraphics[angle=0,width=15cm]{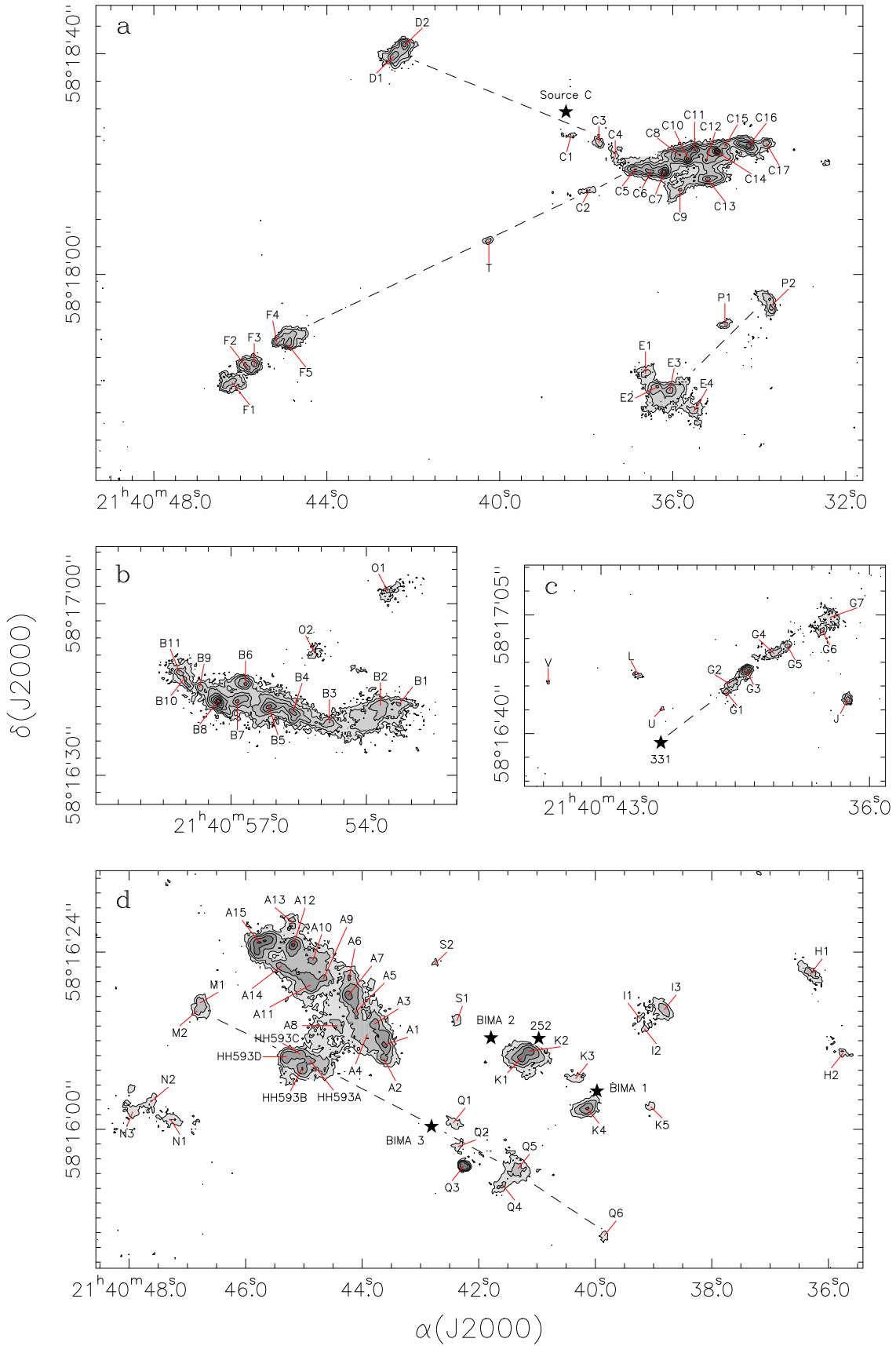}}
\caption{H$_2$ (2.12~$\mu$m) close-up images (continuum-subtracted) of the resolved knots in
the jet features towards IC~1396N, including the names of all the knots. 
The contours are in units of $5.7 \times 10^{-16}$ erg cm$^{-2}$ s$^{-1}$ arcsec$^{-2}$
(1, 2, 4, 8, 13, 20, 28 and 33). 
The black stars show the position of the 1.3~mm
continuum source C detected by Codella et al.~(\cite{codella01}) (top panel), of source \# 331 (middle
right panel), and of source \# 252 and the 3.1~mm sources, BIMA~1,
2, and 3 detected by Beltr\'an et al.~(\cite{beltran02}) (bottom panel). The dashed lines indicate
the orientation of the possible H$_2$ flows described in the text (see Sect.~\ref{H2_emission}). Due to the
complexity of the 
flow powered by
BIMA~2 (see Sect.~\ref{BIMA2_flow}), its direction is not indicated.}
\label{knots_total}
\end{figure*}

Figure~\ref{h2-spitzer} shows the 2.12~$\mu$m H$_2$ integrated line emission in grey-scale
overlaid with an IRAC 4.5~$\mu$m image in color obtained from the Spitzer Center Archive using
the Leopard software. The title of the Spitzer program is {\it Star Formation in Bright Rimmed
Clouds}, and the principal investigator is Giovanni Fazio. The CO (\juz) emission integrated in the low-velocity outflow interval
$[\pm3.5,\pm9.5]~\mbox{km s}^{-1}$ is also shown in red and blue contours (Beltr\'an et al.\
\cite{beltran02}). Figure~\ref{knots_total} shows a close-up of the H$_2$ emission line
features. As seen in these figures, the H$_2$ emission shows a large number of molecular
hydrogen features spread over the region. Many of these emission features have already been
found by Nisini et al.~(\cite{nisini01}), Reipurth et al.~(\cite{reipurth03}), and Caratti o
Garatti~(\cite{caratti06}). We have continued and expanded, when new H$_2$ features have been
discovered, the nomenclature started by these authors (see Table~\ref{knots}). Most of the H$_2$ features are also
visible in the IRAC 4.5~$\mu$m image (Fig.~\ref{h2-spitzer}). The fact that the IRAC
4.5~$\mu$m band is very efficient in detecting Herbig-Haro object is due to the the fact that
the spectral response function is highest in this band, and that between 4--5~$\mu$m there are
many vibrational and rotational H$_2$ emission lines (see, e.\ g.\ Smith \& Rosen~\cite{smith05}).
In addition, the  4.5~$\mu$m band is the less affected by polycyclic aromatic hydrocarbons
(PAHs), whose emission could hide the shock excited H$_2$ features of the HH flows.

The H$_2$ emission is not smooth but complex and knotty, with several condensations embedded in
a more diffuse and nebular emission (Fig.~\ref{knots_total}). The deep and sub-arcsecond
resolution H$_2$ observations have allowed us to resolve the emission into several chains of
knots that could be tracing different flows. Particularly interesting are the chains of knots 
labeled A, B, and C, for which more than 10 individual knots have been mapped. The peak position
and photometry of the individuals knots are given in Table~\ref{knots}. We have named the
individual components of the different H$_2$ emission line features by numbers. These chains or
groups of H$_2$ knots show sometimes a jet-like morphology which together with the fact that 
are located in  different parts of the globule, not just on the bright rim, suggests that the
H$_2$ excitation is mostly due to shocks driven by outflows powered by YSOs (Nisini et al.~\cite{nisini01}).

\subsubsection{H$_2$ knots and flows towards the BIMA sources.}
\label{BIMA2_flow}

\begin{figure}
\centerline{\includegraphics[angle=0,width=8.5cm]{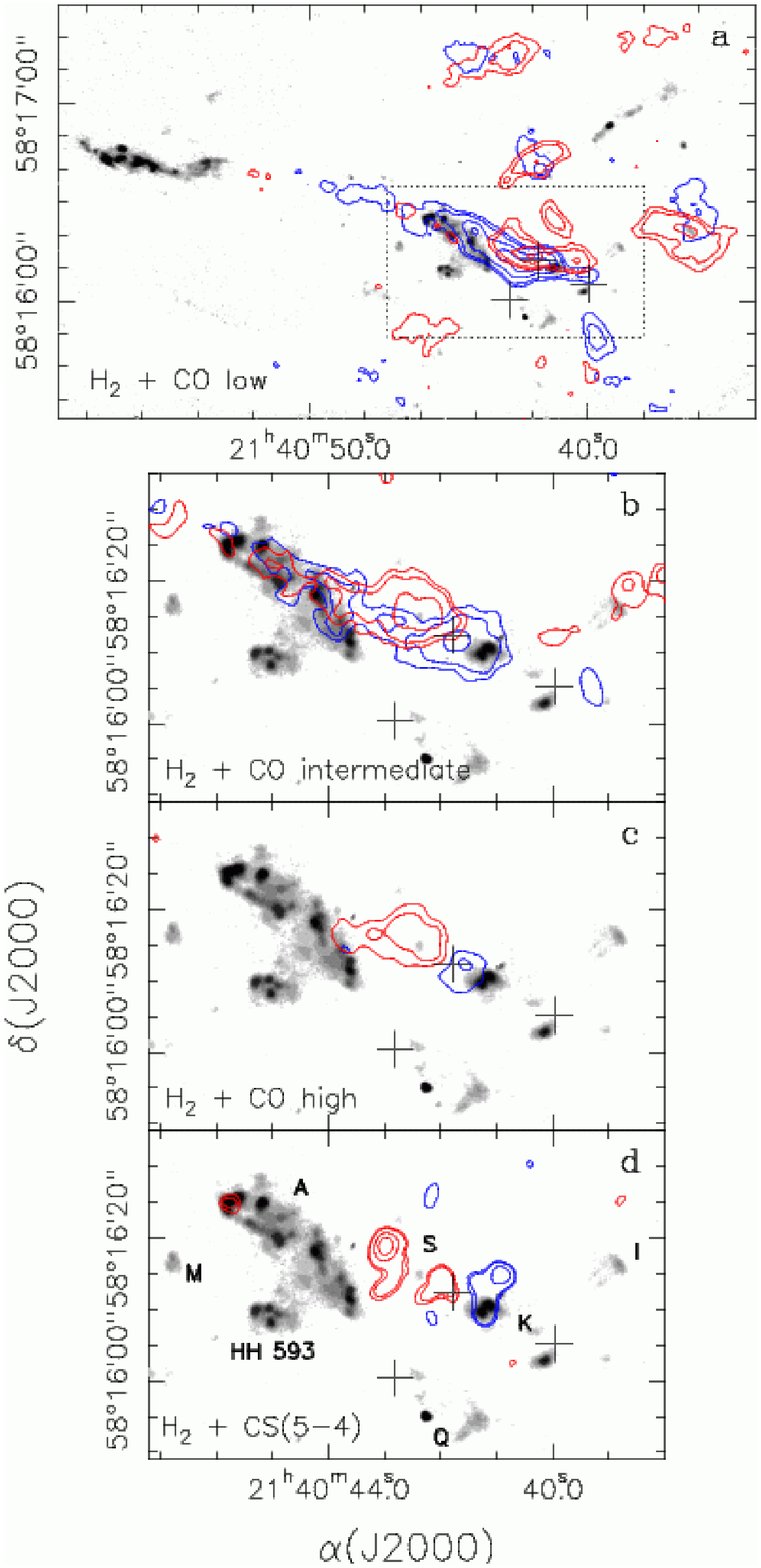}}
\caption{({\it a}): H$_2$ (2.12~$\mu$m) image (continuum-subtracted)  in grey-scale and CO (\juz)
emission integrated in the $[\pm3.5,\pm9.5]~\mbox{km s}^{-1}$ velocity interval in red and 
blue contours (Beltr\'an et al.\ \cite{beltran02}). ({\it Lower panels}): H$_2$ (2.12~$\mu$m) close-up image (continuum-subtracted), towards the 
position of BIMA 1, 2, and 3, in grey-scale and CO (\juz) emission integrated in the
 the $[\pm9.5,\pm15.5]~\mbox{km s}^{-1}$ interval ({\it b}), and the $[\pm15.5,\pm21.5]~\mbox{km s}^{-1}$ interval 
 ({\it c}), and CS
(\jff) emission integrated in the $[\pm5,\pm12]~\mbox{km s}^{-1}$ velocity interval ({\it d}), in red and blue contours (Beltr\'an et al.\ \cite{beltran02}). The black crosses show
the positions of the 3.1~mm sources, BIMA~1, 2, and 3 from Beltr\'an et
al.~(\cite{beltran02}).}
\label{h2-gas}
\end{figure}

Beltr\'an et al.~(\cite{beltran02}) observed the region surrounding the intermediate-mass YSO
\iras\ (BIMA~2) in several molecular tracers and continuum at millimeter wavelengths with the BIMA
interferometer. These authors resolved the millimeter continuum emission into three sources,
BIMA~1, 2, and 3, and mapped in CO two molecular outflows: a north-south outflow powered by
BIMA~1, and an east-west one driven by BIMA~2. The latter outflow shows a very complex morphology
and kinematics because while at high outflow velocities the outflow is clearly bipolar, with the
blueshifted emission towards the west and the redshifted one towards the east, at low outflow
velocities the direction of the outflow gets deflected and the blueshifted and redshifted emission are highly overlapped (see Fig.~\ref{h2-gas}).
Beltr\'an et al.~(\cite{beltran02}) explain the complexity of this outflow in terms of a shocked
cloudlet model scenario, in which the molecular outflow would interact with the dense material
surrounding the embedded sources. The outflow is almost on the plane of the sky (Codella et
al.~(\cite{codella01}) assume an inclination angle of 10\degr--20\degr), which would explain why the
red- and blueshifted outflow emission overlap after the shock. In this section, we want to study in more detail the morphology
of the BIMA~2 outflow and check the validity of the shocked scenario by comparing the 2.12~$\mu$m
H$_2$ emission with that of the outflow as seen in CO and CS with the BIMA interferometer. The top
panel of Fig.~\ref{h2-gas} shows in red and blue contours the CO (\juz) emission integrated in the
low-velocity outflow interval (Beltr\'an et al.~\cite{beltran02}) overlapped on the 2.12~$\mu$m
H$_2$ integrated line emission in grey-scale. The angular resolution of this image is much higher
than that of Fig.~5 of Nisini et al.~(\cite{nisini01}). What is more, the interferometer has
filtered out the extended emission of the outflow allowing to better study the innermost part of
the outflow and to compare the correlation between the H$_2$ and the CO emission. The bottom
panels of Fig.~\ref{h2-gas} show a close-up image of the H$_2$ emission towards the embedded YSOs
BIMA~1, 2, and 3,  overlaid with the CO (\juz) emission integrated in the intermediate-
(Fig.~\ref{h2-gas}b), and high-velocity outflow interval (Fig.~\ref{h2-gas}c), and the CS (\jff)
integrated emission (Fig.~\ref{h2-gas}d) of the BIMA~2 outflow in red and blue contours (Beltr\'an
et al.\ \cite{beltran02}).

Figure~\ref{h2-gas}c shows that at high outflow velocities, the CO outflow, which has a
well-defined bipolar structure, stops before reaching the position of the H$_2$ emission.
Interestingly, on the west side, the CO emission stops in front of the strong H$_2$ knots K1 and
K2.  The position of these two knots coincides with a blueshifted clump visible in CS (\jff)
(Fig.~\ref{h2-gas}d) and CH$_3$OH (\jff) and identified as clump B by Beltr\'an et
al.~(\cite{beltran02}). These authors suggest that there is a shocked surface at the position of
this clump, which would be the responsible of the deflection and V-shaped morphology of the
molecular outflow at low and intermediate outflow velocities (Fig.~\ref{h2-gas}a, b) westwards of
BIMA~2. The detection of the H$_2$ knots K1 and K2 seems to confirm this scenario.  Note that the
knot K3, the strand of knots I, and the knot H1 are associated with the deflected CO emission
(Fig.~\ref{h2-gas}a, b). In particular, H1 is associated with a redshifted and blueshifted CO
clump,    also visible in CS (\jdu) (Beltr\'an et al.~\cite{beltran02}). Eastwards of BIMA~2,
Beltr\'an et al.~(\cite{beltran02}) suggest that the redshifted clump visible in CS (\jff)
(Fig.~\ref{h2-gas}d) and CH$_3$OH (\jff) and named R is the responsible for the change in the
velocity of the gas in the outflow. The eastern CO emission, between the driving source and clump
R, is mainly redshifted at intermediate  and high outflow velocities, whereas farther away from
clump R the outflow emission is slightly stronger in the blue wing than in the red wing.  This
clump R, however, is not visible in H$_2$ emission.  A possible explanation for this could be that
the H$_2$ molecules are dissociated in the shock, and thus, there is no enough H$_2$ to be
detected. However, the most plausible  explanation would be that the R clump, which is located
towards the redshifted lobe of the outflow, is not visible in H$_2$ due to the extinction produced
by the circumstellar material surrounding BIMA~2. Submillimeter observations carried out with the
James Clerk Maxwell Telescope by Correia (\cite{correia00})  have estimated a circumstellar mass
of $\sim$ 20~$M_\odot$  associated with BIMA~2. There is indeed a small knot labeled S1
(Figs.~\ref{knots_total}d and \ref{h2-gas}d) in the red lobe side of the outflow. Although its
position does not exactly coincide with that of the clump R, we cannot discard the possibility
that this knot is part of the shell of the deflecting clump, where the extinction is lower.  The
H$_2$ emission is clearly detected farther east probably because the emission has shifted from
redshifted to blueshifted and the H$_2$ flow has reached outside the core surrounding BIMA~2. In
fact, the H$_2$ emission of the chain of knots A is clearly associated with the CO emission at 
low and intermediate outflow velocities (Fig.~\ref{h2-gas}a, b). The knot labeled A15
(Fig.~\ref{knots_total}d), which is located at the tip of the chain of knots A is probably a
bow-shock, as suggested by its curved morphology, and by its association with high-density gas as
traced by CS~(\jff) and  CS~(\jdu) (Fig.~\ref{h2-gas} and Beltr\'an et al.~\cite{beltran02}).

The CO emission extends farther out towards the east, in the direction of the chain of
knots B. Unfortunately, this feature is located too far from the phase center of the millimeter
observations, and therefore, the interferometer is not sensitive to the emission. In any case,
Nisini et al.~(\cite{nisini01}) and Codella et al.~(\cite{codella01}) show that there is CO
emission also at the position of the strand of knots B. If the chain of knots A and B were both
associated with the BIMA~2 outflow, then the length of this flow eastwards of BIMA~2 would be of
$\sim$ 2$\farcm$3 or 0.5~pc at a distance of 750~pc, which corresponds to a dynamical timescale,
$t_{\rm dyn}$, of $\sim$ 3300~yr for a typical jet velocity, $v_{\rm jet}$, of 150~\kms\ (Reipurth
\& Bally \cite{reipurth01}). Note, however, that to connect the intermediate-mass protostar BIMA~2
with the knots in A and the knots in B, one cannot follow a straight line but a curved one
(Fig.~\ref{h2-gas}a). Such a curve is also visible in the CO emission and could suggest wiggling
or precession of the flow. The precession of the BIMA~2 outflow could be possible if the powering
source were a binary system instead. As a matter of fact, the intermediate-mass source BIMA~2 is
not associated with a binary system but with a cluster, as recently reported by Neri et
al.~(\cite{neri07}), with at least 3 cores embedded inside a common envelope.

Regarding the other knots in the region,  it is possible to trace a straight line to connect
BIMA~3 with the group of knots labeled Q towards the southwest, and with the knots HH593  and the
knots M1 and M2  towards the northeast. A dashed line in Fig.~\ref{knots_total}d indicates the
possible direction of this flow. The straight line can be extrapolated up to reach the group of
knots labeled B. In particular, the line can be traced towards knots B1 and B2 that have a
different orientation from the rest of knots in B. This, together with the  bow-shock
shape of knots Q4 and Q5 (Fig.~\ref{knots_total}d), makes us to speculate with the possibility 
that BIMA~3 is driving an additional flow, which could have a length from BIMA~3 to B1 and B2 of
$\sim$ 1$\farcm$6 or 0.35~pc, and $t_{\rm dyn}$$\simeq$ 2300~yr for $v_{\rm jet}$=150~\kms. The
length of the flow from  Q6 to M1 and M2 is $\sim$ 1$\farcm$1 or 0.24~pc. Although no CO~(\juz) emission has been detected towards the position of this possible
outflow through interferometric observations (Fig.~\ref{h2-gas}), Codella et
al.~(\cite{codella01}) have observed CO~(\jdu) through single-dish observations. Therefore, this
could be an old and poor collimated outflow, whose emission has been filtered out by the
interferometer. 

No H$_2$ emission has been clearly found in association with the north-south outflow mapped in
CO~(\juz) by  Beltr\'an et al.~(\cite{beltran02}) and proposed to be driven by BIMA~1, nor with
a possible additional CS~(\jdu) outflow observed westwards of the BIMA~1 outflow (Beltr\'an et
al.~\cite{beltran04}).

\subsubsection{Other possible H$_2$ flows in the globule}

One of the H$_2$ features mapped in the region by Nisini et al.~(\cite{nisini01}) is the  chain of
knots labeled G. As seen in Fig.~\ref{knots_total}c, this strand of knots shows a jet-like
morphology, with an elongation of $\sim$ $0\farcm75$ or $\sim$ 0.16~pc. This possible flow  seems to
be emanating from source  \# 331, named  HH777 IRS by Reipurth et al.~(\cite{reipurth03}), which
as discussed in Sect.~\ref{population} is possibly binary. One of the sources in this binary system could
be driving the HH777 flow mapped by Reipurth et al.~(\cite{reipurth03}). The 
HH777 flow has a PA of about $-$120\degr. This flow is visible in
H$_\alpha$ and [SII] as a bright working surface abruptly emerging from the southwestern sharp rim
of the cloud core (Reipurth et al.~\cite{reipurth03}). The bow-shock is also visible in the
Digital Sky Survey 2 optical image (see Fig.~1 of Beltr\'an et al.~\cite{beltran02}). Our H$_2$
observations do not cover the position of this bow-shock. As for the HH777 flow before emerging
from the rim, that is, closer to source \# 331, no H$_2$ emission has been detected. Regarding the
chain of knots G, it is not easily associated with any of the CO~(\jdu) peaks mapped by Codella et
al.~(\cite{codella01}), although it is clearly located in a region of enhanced high-velocity CO
emission (see Fig.~5 of Nisini et al.~\cite{nisini01}). No CO~(\juz) emission is detected either
in association with the G knots in the interferometric maps of Beltr\'an et
al.~(\cite{beltran02})  (see Fig.~\ref{h2-spitzer}). This could be due to the fact that this
feature is located too far from the phase center of the millimeter observations, and therefore,
the interferometer is not sensitive to the emission. Small blueshifted and redshifted CO emission
clumps have been detected towards the position of \# 331 (Fig.~\ref{h2-spitzer}), and Caratti o
Garatti et al.~(\cite{caratti06}) have detected [FeII] as well. The dynamical timescale of the
possible G flow would be of $\sim$ 1000~yr for a typical $v_{\rm jet}$. Interestingly, the strand of
knots G points towards a northwestern cavity clearly visible in the IRAC~3.6, 4.5, 5.8, and
8.0~$\mu$m images, which makes us to speculate with the possibility that such a cavity has been
excavated by a molecular outflow that could be associated with these H$_2$ knots.

Towards the north of the IC~1396N globule there are some other prominent groups or chains of
knots, labeled C, D, E, F, and P. Nisini et al.~(\cite{nisini01}) propose that the strands of
knots E and F could be associated with a flow, which they called outflow II, and that the chains
of knots C and D could be associated with the two lobes of a same flow, called outflow III. In the
latter scenario, the powering source of the possible C--D flow would be the 1.3~mm continuum
source C detected by Codella et al.~(\cite{codella01}) and visible in the top panel of
Fig.~\ref{knots_total}. However, as seen in Fig.~\ref{knots_total}a, the group of knots E seems to
be more likely associated with those labeled P, with the knot P2 showing a sort of bow-shock
morphology.  A dashed line in Fig.~\ref{knots_total}a indicates the possible direction of this
flow. Regarding the flow C--D, there are some individual knots in the chain of knots C (C1, C3,
and C4) expanding towards the group of knots D, as one would expect in the scenario proposed by
Nisini et al.~(\cite{nisini01}). The length of this flow would be $\sim$ $0\farcm84$ or
$\sim$ 0.18~pc. If the 1.3~mm continuum source C is powering the flow  C--D, then its $t_{\rm dyn}$
would be $\sim$ 750~yr for a typical $v_{\rm jet}$. It should be noted, however, that there are
also some knots in C expanding towards the chain of knots  F, and there is an additional knot,
labeled T, located between the knots chains F and C. This suggests a possible  association of the
strands of knots C, T, and F that could be part of the same flow (see Fig.~\ref{knots_total}a). The
length of this possible flow would be $\sim$ $1\farcm8$ or $\sim$ 0.40~pc. In fact, Spizer
4.5~$\mu$m observations seem  to give support to this scenario, as the infrared emission connects
F, T, and C (see Fig.~\ref{h2-spitzer}). The two strands of knots C and F could be associated with
the redshifted and blueshifted lobes, respectively, of the northern outflow orientated east-west
mapped in CO~(\jdu) by Codella et al.~(\cite{codella01}). Note that in this case, the redshifted
lobe would be that of the outflow II, according to Nisini et al.~(\cite{nisini01}), while the
blueshifted lobe would be that of the outflow III. Regarding the source powering this flow, it
could be embedded in the dense gas detected in CS by Codella et al.~(\cite{codella01}) and
H$^{13}$CO$^+$ by Sugitani et al.~(\cite{sugitani02a}). An emission peak,  labeled as Core I, is
visible in the Sugitani et al.\ maps. The high-density gas emission coincides with a high
extincted elongated region clearly visible in the J and H band maps (see Fig.~\ref{fig:kmap}). 
Therefore, a possible scenario could be the presence of two flows, one of them traced by the
strands of knots F, T, and some knots of the C strand, and the other one by the group of knots D
and some other individual knots of C. These two flows would collide towards the position of the
chain of knots C, which would explain the strong H$_2$ emission towards this feature.  Note that
by extrapolating southwards the line that connects the chains of knots  C, T, and F, one finds the
group of knots labeled O, and by continuing farther south, the chain of knots B. Therefore, one
could hypothesize that all these groups of knots could be related and be part of a long chain of
H$_2$ emission knots. The total length of this long H$_2$ flow would be of $\sim$ $3\farcm1$ or
$\sim$ 0.70~pc at the distance of IC~1396N.

From the morphology only it is difficult to confirm the possible flows observed in IC~1396N. To
study the kinematics and physical conditions of the H$_2$  emission and determine whether
different H$_2$ features that seem to be morphologically  related are indeed part of the same long
flow, additional long-slit NIR spectroscopy observations would be needed.  In
addition, if $v_{\rm jet}$$\simeq$ 150~\kms, then in about  $\sim$ 5~yr we should be able to
cross-correlate the images and measure displacements of the knots of the order of the pixel size of
NICS at the TNG. This way one could determine proper motions that would help to confirm or discard
possible flows.

\section{Summary and conclusions}

We have carried out deep NIR observations at $J, H$ and $K'$ with NICS at the TNG telescope 
to study in detail the stellar population associated with the bright-rimmed cloud IC~1396N  and
reveal the presence of additional young sources inside the globule.  The deep high angular
resolution H$_2$ observations helped us to investigate the complex structure of this globule, and
the morphology of the shocked gas that traces the interaction between the outflows in the region
and the dense clumps surrounding the YSOs. 


We have found 736 sources detected in all three bands within the area where the $JHK'$ images
overlap  ($\sim$ $4.2\times4.2$ arcmin$^{2}$). There are 128 sources detected only in $HK'$, 67 detected only in $K'$, and 79 detected
only in $JH$. The  completeness limits in the 2MASS standard
are $K_{s} \sim$ 17.5, $H \sim$ 18.5 and $J \sim$ 19.5. The
sources with $HK'$ or $K'$ detections only are preferentially located towards the globule, whereas
the sources with $JH$ detections tend to be located outside the globule. Most of the stars in the
region either fall within the reddening band of the main sequence or exhibit only
a small NIR excess as
shown by the CCD. The source  \# 331, which coincides with source \# 8 in Nisini et
al.~(\cite{nisini01}) and HH777 IRS in Reipurth et al.~(\cite{reipurth03}), is possibly a binary and the
photometry has been derived for both A and B components. The sources of this system could be
powering the major flow HH777 (Reipurth et al.~(\cite{reipurth03}) and the H$_2$ flow G (Nisini et
al.~\cite{nisini01}).  Although there are signatures of star formation in the globule, such
as molecular outflows and jets, only the source \# 331A exhibits a large NIR excess. This and
source \# 252 have been tentatively classified as Class~I sources of intermediate-mass based on the CMD. 

We have not found any color or age gradient in the south-north direction, indicative of the
triggered star formation scenario suggested by Getman et al.~(\cite{getman07}) from X-ray
observations. We have not found either clear evidence of clustering of sources with NIR excess
towards the southern edge of the globule. The evolutionary gradient found by Getman et 
al.~(\cite{getman07}) may not correspond to an age gradient, since the intense UV radiation may
have affected the circumstellar environments of the protostars close to the rim, suddenly stopping
their growth and making them appear as less evolved Class II/III sources. Anyway,  what is clear
from NIR and millimeter observations is that not all the star formation in the globule can be
explained in terms of triggering.

The H$_2$ emission is complex and knotty and shows a large number of molecular hydrogen features
spread over the region, testifying a recent star-formation activity throughout the whole globule.
The H$_2$ emission is resolved into several chains or groups of knots that sometimes show a
jet-like morphology. This together with the fact that the knots are located in  different parts of
the globule suggest that the H$_2$ excitation is mostly due to shocks driven by outflows powered
by YSOs. The shocked cloudlet model scenario proposed by Beltr\'an et al.~(\cite{beltran02}) to
explain the V-shaped morphology of the CO molecular outflow  powered by BIMA~2 seems to be
confirmed by the presence of H$_2$ emission (knots K1 and K2) at the position of the western clump
B, which is causing the deflection of the outflow. The eastern deflecting clump R, visible in
high-density tracers, is not visible in H$_2$, but this could be due to extinction. The H$_2$
emission of this BIMA~2 flow is visible farther east associated with the chain of knots A,
probably when the emission reaches outside the core surrounding BIMA~2. 

New possible flows have been discovered in the globule. One of them would be denoted by the group
of knots E and those labeled P. Another flow would be traced by the strands of knots F, T, and
some knots of the C strand. This flow would collide, towards the position of the chain of knots C,
with the previously known flow C--D, denoted by the group of knots D and some individuals knots of
C. The C--T--F flow could extend farther southwards up to reach the group of knots O or even the
chain of knots B. Another possible flow has also been discovered towards the south of the globule
that would be traced by the group of knots Q, HH593 and M, and could be powered by the YSO BIMA~3.
This flow could extend up to reach some knots of the strand of knots B.  In order to confirm these
flows, additional long-slit NIR spectroscopy observations and proper motions
determination would be needed.

\begin{acknowledgements}

This work is
based on observations made with the Italian Telescopio Nazionale Galileo
(TNG) operated on the island of La Palma by the Fundaci\'{o}n Galileo Galilei of
the INAF (Istituto Nazionale di Astrofisica) at the Spanish Observatorio del
Roque de los Muchachos of the Instituto de Astrofisica de Canarias.

This work is based in part on observations made with the Spitzer Space Telescope,
which is operated by the Jet Propulsion Laboratory, California Institute of Technology
under a contract with NASA.

This publication makes use of data products from the Two Micron All Sky
Survey, which is a  joint project of the University of Massachusetts and the
Infrared Processing and Analysis Center/California Institute of Technology,
funded by the National Aeronautics and Space Administration and the National
Science Foundation.

MTB, RL, JMG, and RE are supported by MEC grant AYA2005-08523-C03. JMG is also supported by AGAUR
grant 2005SGR00489. FM acknowledges support from the Universitat de Barcelona during the data
calibration process.  This publication makes use of data products from the Two Micron All Sky
Survey, which is a joint project of the University of Massachusetts and the  Infrared Processing
and Analysis Center/California Institute of Technology, funded by the National Aeronautics and
Space Administration and the National Science Foundation.

 \end{acknowledgements}

\Online

\begin{table*}
\caption[] {$JHK_s$ photometry and positions of found $K_s$ sources detected in the three bands towards IC 1396N.
}
\label{fot_jhk}
\begin{tabular}{lccccc}
\hline
\multicolumn{1}{c}{NIR} &
\multicolumn{1}{c}{$\alpha({\rm J2000})$} &
\multicolumn{1}{c}{$\delta({\rm J2000})$} &
\multicolumn{1}{c}{$J$}&
\multicolumn{1}{c}{$H$}&
\multicolumn{1}{c}{$K_s$} \\
\multicolumn{1}{c}{source} &
\multicolumn{1}{c}{($^{\rm h}$ $^{\rm m}$ $^{\rm s}$)} &
\multicolumn{1}{c}{($\degr$ $\arcmin$ $\arcsec$)} &
&\multicolumn{1}{c}{(mag)}& \\
\hline
  1 & 21 40 27.76 &58 14 54.33   &   17.732   $\pm$  0.014 &16.970  $\pm$   0.011 &  17.874  $\pm$   0.126  \\
  2 & 21 40 43.25 &58 14 54.51   &   17.788   $\pm$  0.022 &16.972  $\pm$   0.016 &  17.681  $\pm$   0.110  \\
  3 & 21 40 39.10 &58 14 54.69   &   17.598   $\pm$  0.012 &16.938  $\pm$   0.009 &  17.201  $\pm$   0.076  \\
  5 & 21 40 42.12 &58 14 54.82   &   17.419   $\pm$  0.013 &16.523  $\pm$   0.009 &  16.439  $\pm$   0.037  \\
  7 & 21 40 43.29 &58 14 55.57   &   17.269   $\pm$  0.011 &16.590  $\pm$   0.009 &  16.370  $\pm$   0.028  \\
  8 & 21 40 32.01 &58 14 56.29   &   16.370   $\pm$  0.005 &15.726  $\pm$   0.004 &  15.601  $\pm$   0.015  \\
\hline					   
\end{tabular}\end{table*}		   
\addtocounter{table}{-1}		   
					   
\begin{table*}				   
\caption[] {Continued.} 		   
\begin{tabular}{lccccc}
\hline
\multicolumn{1}{c}{NIR} &
\multicolumn{1}{c}{$\alpha({\rm J2000})$} &
\multicolumn{1}{c}{$\delta({\rm J2000})$} &
\multicolumn{1}{c}{$J$}&
\multicolumn{1}{c}{$H$}&
\multicolumn{1}{c}{$K_s$} \\
\multicolumn{1}{c}{source} &
\multicolumn{1}{c}{($^{\rm h}$ $^{\rm m}$ $^{\rm s}$)} &
\multicolumn{1}{c}{($\degr$ $\arcmin$ $\arcsec$)} &
&\multicolumn{1}{c}{(mag)}& \\
\hline
   93 & 21 40 47.61 &58 15 20.89 &   18.568   $\pm$  0.031     &17.716  $\pm$	0.020 &17.308  $\pm$   0.040\\
   94 & 21 40 32.79 &58 15 21.68 &   17.063   $\pm$  0.006     &16.098  $\pm$	0.003 &15.788  $\pm$   0.011\\
   96 & 21 40 41.28 &58 15 22.01 &   18.986   $\pm$  0.050     &17.866  $\pm$	0.029 &17.182  $\pm$   0.043\\
   98 & 21 40 51.08 &58 15 21.79 &   17.757   $\pm$  0.014     &16.911  $\pm$	0.010 &16.697  $\pm$   0.030\\
   99 & 21 40 50.63 &58 15 22.22 &   15.277   $\pm$  0.002     &14.836  $\pm$	0.002 &14.591  $\pm$   0.003\\
  100 & 21 40 27.24 &58 15 23.03 &   18.002   $\pm$  0.016     &17.268  $\pm$	0.010 &17.013  $\pm$   0.043\\
\hline					   
\end{tabular}\end{table*}		   
\addtocounter{table}{-1}		   
					   
\begin{table*}				   
\caption[] {Continued.} 		   
\begin{tabular}{lccccc}
\hline
\multicolumn{1}{c}{NIR} &
\multicolumn{1}{c}{$\alpha({\rm J2000})$} &
\multicolumn{1}{c}{$\delta({\rm J2000})$} &
\multicolumn{1}{c}{$J$}&
\multicolumn{1}{c}{$H$}&
\multicolumn{1}{c}{$K_s$} \\
\multicolumn{1}{c}{source} &
\multicolumn{1}{c}{($^{\rm h}$ $^{\rm m}$ $^{\rm s}$)} &
\multicolumn{1}{c}{($\degr$ $\arcmin$ $\arcsec$)} &
&\multicolumn{1}{c}{(mag)}& \\
\hline
  225 & 21 40 27.38 &58 16 00.23 &   17.886   $\pm$  0.013 &17.200  $\pm$   0.009 &16.875  $\pm$   0.030    \\
  227 & 21 40 48.60 &58 16 01.06 &   15.575   $\pm$  0.002 &13.305  $\pm$   0.001 &11.843  $\pm$   0.001    \\
  229 & 21 40 31.59 &58 16 02.11 &   18.084   $\pm$  0.017 &16.858  $\pm$   0.007 &16.185  $\pm$   0.012    \\
  230 & 21 40 28.11 &58 16 02.48 &   18.662   $\pm$  0.023 &17.865  $\pm$   0.019 &17.544  $\pm$   0.054    \\
  232 & 21 40 45.51 &58 16 02.81 &   15.712   $\pm$  0.003 &13.795  $\pm$   0.001 &12.829  $\pm$   0.001    \\
  233 & 21 40 32.48 &58 16 03.79 &   18.908   $\pm$  0.036 &17.045  $\pm$   0.008 &16.059  $\pm$   0.011    \\
\hline					   
\end{tabular}\end{table*}		   
\addtocounter{table}{-1}		   
					   
\begin{table*}				   
\caption[] {Continued.} 		   
\begin{tabular}{lccccc}
\hline
\multicolumn{1}{c}{NIR} &
\multicolumn{1}{c}{$\alpha({\rm J2000})$} &
\multicolumn{1}{c}{$\delta({\rm J2000})$} &
\multicolumn{1}{c}{$J$}&
\multicolumn{1}{c}{$H$}&
\multicolumn{1}{c}{$K_s$} \\
\multicolumn{1}{c}{source} &
\multicolumn{1}{c}{($^{\rm h}$ $^{\rm m}$ $^{\rm s}$)} &
\multicolumn{1}{c}{($\degr$ $\arcmin$ $\arcsec$)} &
&\multicolumn{1}{c}{(mag)}& \\
\hline
  339 & 21 40 29.13 &58 16 39.99 &   18.151   $\pm$  0.031  &17.108  $\pm$   0.018 &  16.651  $\pm$   0.034 \\
  340 & 21 40 22.32 &58 16 40.09 &   18.418   $\pm$  0.024  &17.734  $\pm$   0.021 &  17.420  $\pm$   0.054 \\
  342 & 21 40 30.15 &58 16 40.43 &   18.547   $\pm$  0.030  &17.500  $\pm$   0.014 &  16.922  $\pm$   0.028 \\
  343 & 21 40 22.94 &58 16 41.34 &   18.080   $\pm$  0.017  &17.296  $\pm$   0.015 &  16.988  $\pm$   0.031 \\
  345 & 21 40 27.70 &58 16 41.51 &   17.440   $\pm$  0.016  &16.625  $\pm$   0.011 &  16.219  $\pm$   0.027 \\
  347 & 21 40 45.25 &58 16 41.59 &   17.376   $\pm$  0.011  &16.296  $\pm$   0.006 &  15.531  $\pm$   0.007 \\
\hline					   
\end{tabular}\end{table*}		   
\addtocounter{table}{-1}		   
					   
\begin{table*}				   
\caption[] {Continued.} 		   
\begin{tabular}{lccccc}
\hline
\multicolumn{1}{c}{NIR} &
\multicolumn{1}{c}{$\alpha({\rm J2000})$} &
\multicolumn{1}{c}{$\delta({\rm J2000})$} &
\multicolumn{1}{c}{$J$}&
\multicolumn{1}{c}{$H$}&
\multicolumn{1}{c}{$K_s$} \\
\multicolumn{1}{c}{source} &
\multicolumn{1}{c}{($^{\rm h}$ $^{\rm m}$ $^{\rm s}$)} &
\multicolumn{1}{c}{($\degr$ $\arcmin$ $\arcsec$)} &
&\multicolumn{1}{c}{(mag)}& \\
\hline
  468 & 21 40 25.08 &58 17 18.87 &   18.042   $\pm$  0.015    &17.823  $\pm$   0.021 &17.366  $\pm$   0.051 \\
  469 & 21 40 23.21 &58 17 19.04 &   18.418   $\pm$  0.021    &17.791  $\pm$   0.020 &17.517  $\pm$   0.057 \\
  472 & 21 40 25.14 &58 17 20.06 &   17.940   $\pm$  0.013    &17.326  $\pm$   0.011 &17.101  $\pm$   0.043 \\
  473 & 21 40 49.06 &58 17 19.77 &   15.582   $\pm$  0.002    &13.782  $\pm$   0.001 &12.938  $\pm$   0.001 \\
  475 & 21 40 27.11 &58 17 20.82 &   17.823   $\pm$  0.012    &17.202  $\pm$   0.009 &16.908  $\pm$   0.035 \\
  477 & 21 40 26.86 &58 17 21.33 &   18.890   $\pm$  0.033    &18.203  $\pm$   0.027 &17.810  $\pm$   0.072 \\
\hline					   
\end{tabular}\end{table*}		   
\addtocounter{table}{-1}		   
					   
\begin{table*}				   
\caption[] {Continued.} 		   
\begin{tabular}{lccccc}
\hline
\multicolumn{1}{c}{NIR} &
\multicolumn{1}{c}{$\alpha({\rm J2000})$} &
\multicolumn{1}{c}{$\delta({\rm J2000})$} &
\multicolumn{1}{c}{$J$}&
\multicolumn{1}{c}{$H$}&
\multicolumn{1}{c}{$K_s$} \\
\multicolumn{1}{c}{source} &
\multicolumn{1}{c}{($^{\rm h}$ $^{\rm m}$ $^{\rm s}$)} &
\multicolumn{1}{c}{($\degr$ $\arcmin$ $\arcsec$)} &
&\multicolumn{1}{c}{(mag)}& \\
\hline
  582 & 21 40 21.84 &58 18 02.63 &   17.892   $\pm$  0.013    &17.176  $\pm$   0.013 &16.870  $\pm$   0.031 \\
  584 & 21 40 19.93 &58 18 04.64 &   19.291   $\pm$  0.082    &18.537  $\pm$   0.065 &18.160  $\pm$   0.094 \\
  587 & 21 40 29.13 &58 18 04.83 &   18.675   $\pm$  0.032    &17.586  $\pm$   0.016 &17.150  $\pm$   0.032 \\
  590 & 21 40 42.97 &58 18 05.73 &   19.980   $\pm$  0.112    &18.416  $\pm$   0.033 &17.595  $\pm$   0.046 \\
  591 & 21 40 24.27 &58 18 06.18 &   19.499   $\pm$  0.058    &18.459  $\pm$   0.044 &18.042  $\pm$   0.080 \\
  592 & 21 40 48.11 &58 18 05.76 &   19.177   $\pm$  0.049    &17.831  $\pm$   0.018 &17.197  $\pm$   0.030 \\
\hline					   
\end{tabular}\end{table*}		   
\addtocounter{table}{-1}		   
					   
\begin{table*}				   
\caption[] {Continued.} 		   
\begin{tabular}{lccccc}
\hline
\multicolumn{1}{c}{NIR} &
\multicolumn{1}{c}{$\alpha({\rm J2000})$} &
\multicolumn{1}{c}{$\delta({\rm J2000})$} &
\multicolumn{1}{c}{$J$}&
\multicolumn{1}{c}{$H$}&
\multicolumn{1}{c}{$K_s$} \\
\multicolumn{1}{c}{source} &
\multicolumn{1}{c}{($^{\rm h}$ $^{\rm m}$ $^{\rm s}$)} &
\multicolumn{1}{c}{($\degr$ $\arcmin$ $\arcsec$)} &
&\multicolumn{1}{c}{(mag)}& \\
\hline
  684 & 21 40 46.61 &58 18 33.91 &   18.499   $\pm$  0.027   &16.861  $\pm$   0.007 &15.992  $\pm$   0.011  \\
  686 & 21 40 39.86 &58 18 35.03 &   18.324   $\pm$  0.021   &15.116  $\pm$   0.002 &13.288  $\pm$   0.001  \\
  688 & 21 40 27.01 &58 18 35.49 &   19.881   $\pm$  0.090   &18.465  $\pm$   0.031 &17.759  $\pm$   0.063  \\
  690 & 21 40 30.42 &58 18 36.30 &   18.993   $\pm$  0.041   &17.644  $\pm$   0.018 &17.142  $\pm$   0.024  \\
  691 & 21 40 48.48 &58 18 36.48 &   18.372   $\pm$  0.024   &17.051  $\pm$   0.009 &16.367  $\pm$   0.014  \\
  693 & 21 40 23.08 &58 18 37.52 &   16.618   $\pm$  0.005   &15.893  $\pm$   0.004 &15.553  $\pm$   0.009  \\
\hline					   
\end{tabular}\end{table*}		   
\addtocounter{table}{-1}		   
					   
\begin{table*}				   
\caption[] {Continued.} 		   
\begin{tabular}{lccccc}
\hline
\multicolumn{1}{c}{NIR} &
\multicolumn{1}{c}{$\alpha({\rm J2000})$} &
\multicolumn{1}{c}{$\delta({\rm J2000})$} &
\multicolumn{1}{c}{$J$}&
\multicolumn{1}{c}{$H$}&
\multicolumn{1}{c}{$K_s$} \\
\multicolumn{1}{c}{source} &
\multicolumn{1}{c}{($^{\rm h}$ $^{\rm m}$ $^{\rm s}$)} &
\multicolumn{1}{c}{($\degr$ $\arcmin$ $\arcsec$)} &
&\multicolumn{1}{c}{(mag)}& \\
\hline
  907 & 21 40 36.86 &58 15 22.32 &   19.600   $\pm$  0.082   &18.601  $\pm$   0.049 &18.152  $\pm$   0.120  \\
  908 & 21 40 33.54 &58 15 23.66 &   18.459   $\pm$  0.026   &17.785  $\pm$   0.018 &17.662  $\pm$   0.068  \\
  909 & 21 40 33.22 &58 15 36.32 &   19.442   $\pm$  0.061   &18.217  $\pm$   0.029 &17.984  $\pm$   0.076  \\
  910 & 21 40 36.45 &58 15 03.67 &   18.261   $\pm$  0.026   &17.698  $\pm$   0.025 &18.250  $\pm$   0.092  \\
  911 & 21 40 38.98 &58 15 09.49 &   19.246   $\pm$  0.048   &18.224  $\pm$   0.032 &18.105  $\pm$   0.075  \\
  912 & 21 40 40.00 &58 15 12.97 &   19.436   $\pm$  0.055   &18.228  $\pm$   0.029 &17.889  $\pm$   0.065  \\
\hline					   
\end{tabular}\end{table*}		   
\addtocounter{table}{-1}		   
					   
\begin{table*}				   
\caption[] {Continued.} 		   
\begin{tabular}{lccccc}
\hline
\multicolumn{1}{c}{NIR} &
\multicolumn{1}{c}{$\alpha({\rm J2000})$} &
\multicolumn{1}{c}{$\delta({\rm J2000})$} &
\multicolumn{1}{c}{$J$}&
\multicolumn{1}{c}{$H$}&
\multicolumn{1}{c}{$K_s$} \\
\multicolumn{1}{c}{source} &
\multicolumn{1}{c}{($^{\rm h}$ $^{\rm m}$ $^{\rm s}$)} &
\multicolumn{1}{c}{($\degr$ $\arcmin$ $\arcsec$)} &
&\multicolumn{1}{c}{(mag)}& \\
\hline
 1005 & 21 40 22.87 &58 16 04.10 &   19.463   $\pm$  0.064   &18.380  $\pm$   0.042  & 17.981  $\pm$   0.085  \\
 1007 & 21 40 23.78 &58 15 45.02 &   19.502   $\pm$  0.068   &18.350  $\pm$   0.042  & 17.924  $\pm$   0.085  \\
 1008 & 21 40 23.51 &58 15 43.79 &   18.868   $\pm$  0.037   &17.958  $\pm$   0.029  & 17.518  $\pm$   0.069  \\
 1009 & 21 40 21.08 &58 15 44.97 &   18.143   $\pm$  0.028   &17.618  $\pm$   0.033  & 16.964  $\pm$   0.040  \\
 1010 & 21 40 22.01 &58 15 48.12 &   19.413   $\pm$  0.059   &18.414  $\pm$   0.042  & 17.866  $\pm$   0.074  \\
 1011 & 21 40 22.26 &58 15 56.23 &   18.917   $\pm$  0.033   &18.028  $\pm$   0.030  & 17.465  $\pm$   0.054  \\
\hline					   
\end{tabular}\end{table*}		   
\addtocounter{table}{-1}		   
					   
\begin{table*}				   
\caption[] {Continued.} 		   
\begin{tabular}{lccccc}
\hline
\multicolumn{1}{c}{NIR} &
\multicolumn{1}{c}{$\alpha({\rm J2000})$} &
\multicolumn{1}{c}{$\delta({\rm J2000})$} &
\multicolumn{1}{c}{$J$}&
\multicolumn{1}{c}{$H$}&
\multicolumn{1}{c}{$K_s$} \\
\multicolumn{1}{c}{source} &
\multicolumn{1}{c}{($^{\rm h}$ $^{\rm m}$ $^{\rm s}$)} &
\multicolumn{1}{c}{($\degr$ $\arcmin$ $\arcsec$)} &
&\multicolumn{1}{c}{(mag)}& \\
\hline
 1232 & 21 40 48.32 &58 18 59.47  &   19.623   $\pm$  0.106    &18.727  $\pm$	0.075 &19.020  $\pm$   0.189\\
 1234 & 21 40 46.75 &58 18 36.71  &   20.284   $\pm$  0.135    &18.904  $\pm$	0.053 &18.207  $\pm$   0.080\\
 1235 & 21 40 48.25 &58 18 34.91  &   20.382   $\pm$  0.156    &19.008  $\pm$	0.059 &18.359  $\pm$   0.100\\
 1237 & 21 40 49.53 &58 18 35.63  &   20.331   $\pm$  0.142    &19.053  $\pm$	0.058 &18.891  $\pm$   0.121\\
 1245 & 21 40 52.25 &58 18 24.89  &   19.084   $\pm$  0.098    &18.316  $\pm$	0.036 &17.797  $\pm$   0.066\\
 1246 & 21 40 50.89 &58 18 29.94  &   20.190   $\pm$  0.123    &18.911  $\pm$	0.056 &18.494  $\pm$   0.097\\
\hline					   
\end{tabular}\end{table*}		   
\addtocounter{table}{-1}		   
					   
\begin{table*}				   
\caption[] {Continued.} 		   
\begin{tabular}{lccccc}
\hline
\multicolumn{1}{c}{NIR} &
\multicolumn{1}{c}{$\alpha({\rm J2000})$} &
\multicolumn{1}{c}{$\delta({\rm J2000})$} &
\multicolumn{1}{c}{$J$}&
\multicolumn{1}{c}{$H$}&
\multicolumn{1}{c}{$K_s$} \\
\multicolumn{1}{c}{source} &
\multicolumn{1}{c}{($^{\rm h}$ $^{\rm m}$ $^{\rm s}$)} &
\multicolumn{1}{c}{($\degr$ $\arcmin$ $\arcsec$)} &
&\multicolumn{1}{c}{(mag)}& \\
\hline
 1386 & 21 40 24.24 &58 18 31.27  &   19.845   $\pm$  0.079    &18.669  $\pm$	0.052 &18.151  $\pm$   0.091\\
 1387 & 21 40 26.53 &58 17 48.63  &   19.360   $\pm$  0.063    &18.414  $\pm$	0.041 &17.865  $\pm$   0.069\\
 1388 & 21 40 28.65 &58 17 36.68  &   19.972   $\pm$  0.093    &18.458  $\pm$	0.032 &17.735  $\pm$   0.074\\
 1389 & 21 40 29.40 &58 17 31.53  &   20.199   $\pm$  0.108    &18.865  $\pm$	0.055 &18.246  $\pm$   0.119\\
 1390 & 21 40 22.61 &58 17 48.71  &   19.914   $\pm$  0.100    &19.087  $\pm$	0.072 &18.613  $\pm$   0.153\\
 1391 & 21 40 22.21 &58 17 47.09  &   19.601   $\pm$  0.067    &18.643  $\pm$	0.048 &18.013  $\pm$   0.091\\
\hline					    
\end{tabular}
\\
 (a) The source \# 331 is double (see Sect.~3.1). The two components have been labeled A and B. 
 The source \# 331A has been detected in the 3 bands and \# 331B in the $H$ and $K_s$ bands only. 
\end{table*}		   

\begin{table*}				   
\caption[] {$JHK_s$ photometry and positions of found $K_s$ sources detected in the $H$ and $K_s$ bands.
}
\label{fot_hk} 		   
\begin{tabular}{lccccc}
\hline
\multicolumn{1}{c}{NIR} &
\multicolumn{1}{c}{$\alpha({\rm J2000})$} &
\multicolumn{1}{c}{$\delta({\rm J2000})$} &
\multicolumn{1}{c}{$J$}&
\multicolumn{1}{c}{$H$}&
\multicolumn{1}{c}{$K_s$} \\
\multicolumn{1}{c}{source} &
\multicolumn{1}{c}{($^{\rm h}$ $^{\rm m}$ $^{\rm s}$)} &
\multicolumn{1}{c}{($\degr$ $\arcmin$ $\arcsec$)} &
&\multicolumn{1}{c}{(mag)}& \\
\hline
   95 & 21 40 44.61 &58 15 21.62  &   $>$20.860      &18.819  $\pm$   0.043 &17.613  $\pm$   0.053\\
  144 & 21 40 38.20 &58 15 33.09  &   $>$20.874      &18.318  $\pm$   0.032 &16.723  $\pm$   0.023\\
  162 & 21 40 45.20 &58 15 36.91  &   $>$20.842      &19.402  $\pm$   0.075 &17.456  $\pm$   0.036\\
  206 & 21 40 46.50 &58 15 51.78  &   $>$20.841      &19.455  $\pm$   0.088 &17.492  $\pm$   0.036\\
  231 & 21 40 50.56 &58 16 02.03  &   $>$20.835      &19.646  $\pm$   0.100 &17.553  $\pm$   0.032\\
  252 & 21 40 40.97 &58 16 12.32  &   $>$20.921      &16.670  $\pm$   0.013 &13.840  $\pm$   0.003\\
\hline					   
\end{tabular}\end{table*}		   
\addtocounter{table}{-1}		   
					   
\begin{table*}				   
\caption[] {Continued.} 		   
\begin{tabular}{lccccc}
\hline
\multicolumn{1}{c}{NIR} &
\multicolumn{1}{c}{$\alpha({\rm J2000})$} &
\multicolumn{1}{c}{$\delta({\rm J2000})$} &
\multicolumn{1}{c}{$J$}&
\multicolumn{1}{c}{$H$}&
\multicolumn{1}{c}{$K_s$} \\
\multicolumn{1}{c}{source} &
\multicolumn{1}{c}{($^{\rm h}$ $^{\rm m}$ $^{\rm s}$)} &
\multicolumn{1}{c}{($\degr$ $\arcmin$ $\arcsec$)} &
&\multicolumn{1}{c}{(mag)}& \\
\hline
 1077 & 21 40 33.21 &58 16 33.94  &   $>$20.844      &19.366  $\pm$   0.076  &18.423  $\pm$   0.101\\
 1078 & 21 40 34.16 &58 17 01.23  &   $>$20.841      &19.504  $\pm$   0.089  &18.960  $\pm$   0.182\\
 1081 & 21 40 35.10 &58 17 02.46  &   $>$20.842      &19.424  $\pm$   0.099  &18.285  $\pm$   0.098\\
 1083 & 21 40 37.67 &58 16 40.27  &   $>$20.836      &19.622  $\pm$   0.122  &17.735  $\pm$   0.045\\
 1085 & 21 40 34.79 &58 16 38.13  &   $>$20.850      &19.150  $\pm$   0.083  &17.617  $\pm$   0.045\\
 1125 & 21 40 37.87 &58 17 19.80  &   $>$20.848      &19.237  $\pm$   0.087  &18.120  $\pm$   0.078\\
\hline				           
\end{tabular}
\\
 (a) The source \# 331 is double (see Sect.~3.1). The two components have been labeled A and B. 
 The source \# 331A has been detected in the 3 bands and \# 331B in the $H$ and $K_s$ bands only. 
\end{table*}	           
					   
\begin{table*}				   
\caption[] {$JHK_s$ photometry and positions of sources only detected in the $K_s$ band.
} 
\label{fot_k} 			   
\begin{tabular}{lccccc}
\hline
\multicolumn{1}{c}{NIR} &
\multicolumn{1}{c}{$\alpha({\rm J2000})$} &
\multicolumn{1}{c}{$\delta({\rm J2000})$} &
\multicolumn{1}{c}{$J$}&
\multicolumn{1}{c}{$H$}&
\multicolumn{1}{c}{$K_s$} \\
\multicolumn{1}{c}{source} &
\multicolumn{1}{c}{($^{\rm h}$ $^{\rm m}$ $^{\rm s}$)} &
\multicolumn{1}{c}{($\degr$ $\arcmin$ $\arcsec$)} &
&\multicolumn{1}{c}{(mag)}& \\
\hline
   33 & 21 40 23.31 &58 15 04.30  &   $>$20.795     &$>$20.086    &$>$19.894 \\
   57 & 21 40 51.67 &58 15 11.49  &   $>$20.795     &$>$19.869    &$>$14.702 \\
  993 & 21 40 23.88 &58 15 09.34  &   $>$20.795     &$>$19.933    &$>$16.242 \\
  123 & 21 40 43.52 &58 15 26.65  &   $>$20.795     &$>$19.997    &$>$17.765 \\
  156 & 21 40 19.80 &58 15 36.81  &   $>$20.795     &$>$19.996    &$>$16.648 \\
  196 & 21 40 45.77 &58 15 49.10  &   $>$20.795     &$>$19.993    &$>$17.680 \\
\hline				           
\end{tabular}\end{table*}	           
				           
\begin{table*}				   
\caption[] {$JHK_s$ photometry and positions of sources detected in the $J$ and $H$ bands.
} 		   
\label{fot_jh} 	
\begin{tabular}{lccccc}
\hline
\multicolumn{1}{c}{NIR} &
\multicolumn{1}{c}{$\alpha({\rm J2000})$} &
\multicolumn{1}{c}{$\delta({\rm J2000})$} &
\multicolumn{1}{c}{$J$}&
\multicolumn{1}{c}{$H$}&
\multicolumn{1}{c}{$K_s$} \\
\multicolumn{1}{c}{source} &
\multicolumn{1}{c}{($^{\rm h}$ $^{\rm m}$ $^{\rm s}$)} &
\multicolumn{1}{c}{($\degr$ $\arcmin$ $\arcsec$)} &
&\multicolumn{1}{c}{(mag)}& \\
\hline
10000 & 21 40 19.77 &58 15 08.89   &   18.607	$\pm$  0.063   &17.999  $\pm$	0.058 &$>$19.179  \\
10000 & 21 40 49.57 &58 15 08.90   &   19.882	$\pm$  0.088   &18.743  $\pm$	0.047 &$>$19.215  \\
10000 & 21 40 20.35 &58 15 11.71   &   19.310	$\pm$  0.077   &18.618  $\pm$	0.066 &$>$19.210  \\
10000 & 21 40 21.59 &58 15 13.07   &   19.021	$\pm$  0.042   &18.302  $\pm$	0.050 &$>$19.194  \\
10000 & 21 40 21.91 &58 15 14.92   &   20.409	$\pm$  0.158   &19.645  $\pm$	0.169 &$>$19.261  \\
10000 & 21 40 23.71 &58 15 28.66   &   19.818	$\pm$  0.112   &19.147  $\pm$	0.100 &$>$19.237  \\
\hline					   					       
\end{tabular}\end{table*}		   
\addtocounter{table}{-1}		   
					   
\begin{table*}				   
\caption[] {Continued.} 		   
\begin{tabular}{lccccc}
\hline
\multicolumn{1}{c}{NIR} &
\multicolumn{1}{c}{$\alpha({\rm J2000})$} &
\multicolumn{1}{c}{$\delta({\rm J2000})$} &
\multicolumn{1}{c}{$J$}&
\multicolumn{1}{c}{$H$}&
\multicolumn{1}{c}{$K_s$} \\
\multicolumn{1}{c}{source} &
\multicolumn{1}{c}{($^{\rm h}$ $^{\rm m}$ $^{\rm s}$)} &
\multicolumn{1}{c}{($\degr$ $\arcmin$ $\arcsec$)} &
&\multicolumn{1}{c}{(mag)}& \\
\hline
10000 & 21 40 26.57 &58 17 42.36  &   20.390   $\pm$  0.191    &19.851  $\pm$	0.184 &$>$19.272 \\
10000 & 21 40 25.71 &58 17 42.25  &   20.969   $\pm$  0.252    &19.927  $\pm$	0.145 &$>$19.275 \\
10000 & 21 40 25.84 &58 17 49.89  &   19.849   $\pm$  0.136    &18.949  $\pm$	0.092 &$>$19.226 \\
10000 & 21 40 24.60 &58 17 45.28  &   20.793   $\pm$  0.202    &19.556  $\pm$	0.115 &$>$19.255 \\
10000 & 21 40 36.40 &58 18 47.87  &   19.999   $\pm$  0.118    &19.454  $\pm$	0.135 &$>$19.252 \\
10000 & 21 40 48.41 &58 18 21.10  &   20.724   $\pm$  0.184    &19.995  $\pm$	0.126 &$>$19.279 \\
10000 & 21 40 21.15 &58 15 10.21  &   20.076   $\pm$  0.106    &19.465  $\pm$	0.127 &$>$19.253 \\
10000 & 21 40 24.14 &58 15 35.19  &   20.013   $\pm$  0.120    &19.721  $\pm$	0.156 &$>$19.267 \\
10000 & 21 40 24.61 &58 15 30.92  &   20.007   $\pm$  0.104    &19.478  $\pm$	0.117 &$>$19.254 \\
10000 & 21 40 22.58 &58 15 33.47  &   20.941   $\pm$  0.296    &20.099  $\pm$	0.237 &$>$19.284 \\
10000 & 21 40 22.46 &58 15 39.73  &   20.338   $\pm$  0.180    &19.454  $\pm$	0.134 &$>$19.251 \\
10000 & 21 40 22.36 &58 16 51.10  &   20.107   $\pm$  0.101    &19.429  $\pm$	0.105 &$>$19.251 \\
\hline					   					       	 
\end{tabular}				   						     
\end{table*}

\begin{table*}				   
\caption[] {H$_2$ photometry and positions of the knots.}
\label{knots} 		   
\begin{tabular}{lccc}
\hline
 &
\multicolumn{1}{c}{$\alpha({\rm J2000})$} &
\multicolumn{1}{c}{$\delta({\rm J2000})$} &
\multicolumn{1}{c}{Flux}
 \\
\multicolumn{1}{c}{Knot} &
\multicolumn{1}{c}{($^{\rm h}$ $^{\rm m}$ $^{\rm s}$)} &
\multicolumn{1}{c}{($\degr$ $\arcmin$ $\arcsec$)} &
\multicolumn{1}{c}{(10$^{-15}$ erg cm$^{-2}$ s$^{-1}$)} \\
\hline
HH593A	             & 21 40 44.895 & 58 16 09.31  & 	 7.2519 $\pm$ 0.120231  \\		 
HH593B	             & 21 40 45.030 & 58 16 08.24  & 	 9.7899 $\pm$ 0.135257  \\
HH593C	             & 21 40 45.063 & 58 16 10.18  & 	 6.9856 $\pm$ 0.122249  \\
HH593D	             & 21 40 45.301 & 58 16 09.84  & 	 9.7167 $\pm$ 0.161095  \\
A1	             & 21 40 43.619 & 58 16 11.47  & 	21.5932 $\pm$ 0.278442  \\
A2                   & 21 40 43.630 & 58 16 09.60  & 	 6.3906 $\pm$ 0.182471	\\
\hline				           	     		
\end{tabular}
\\
 (a) New detection.
\end{table*}	           	     		
\addtocounter{table}{-1}

\begin{table*}				   	     
\caption[] {Continued.} 			     
\begin{tabular}{lccc}				     
\hline						     
 &						     
\multicolumn{1}{c}{$\alpha({\rm J2000})$} &	     
\multicolumn{1}{c}{$\delta({\rm J2000})$} &	     
\multicolumn{1}{c}{Flux}			     
 \\						     
\multicolumn{1}{c}{Knot} &
\multicolumn{1}{c}{($^{\rm h}$ $^{\rm m}$ $^{\rm s}$)} &
\multicolumn{1}{c}{($\degr$ $\arcmin$ $\arcsec$)} &
\multicolumn{1}{c}{(10$^{-15}$ erg cm$^{-2}$ s$^{-1}$)} \\
\hline				
K1	             & 21 40 41.266 & 58 16 09.67 &  17.5391 $\pm$ 0.161546  \\
K2	             & 21 40 41.134 & 58 16 10.60 &  20.6042 $\pm$ 0.132845   \\
K3	             & 21 40 40.335 & 58 16 06.90 &   1.9639 $\pm$ 0.090445  \\
K4	             & 21 40 40.136 & 58 16 02.82 &  12.0642 $\pm$ 0.155566  \\
K5	             & 21:40:39.051 & 58:16:03.05 &   1.4307 $\pm$ 0.084337  \\
L	             & 21 40 42.022 & 58 16 52.23 &   2.3914 $\pm$ 0.116738 \\
\hline	        									 
\end{tabular}   									     
\\
 (a) New detection.
\end{table*}

\end{document}